\documentclass[acmlarge]{acmart}
\usepackage{booktabs} % For formal tables
\usepackage{lmodern}

\usepackage{graphicx}
\usepackage{balance}
\usepackage{comment}

\usepackage[vlined, noend,algo2e]{algorithm2e}
\usepackage{verbatim}
\usepackage{epstopdf}
\usepackage{balance}
\usepackage{comment}
\usepackage{subfigure}
\usepackage{booktabs}
\usepackage{amsmath}
\usepackage{amssymb}
\usepackage{amsfonts}
\usepackage{bm}
\usepackage{colortbl}
\usepackage{tabularx}
\usepackage{color}
\usepackage{epsfig}
\usepackage{xspace}
\usepackage{caption}
\usepackage{graphicx}    % For importing graphics
\usepackage{url}         
\usepackage{algorithm}
\usepackage[noend]{algpseudocode}
\usepackage{placeins}
\usepackage{multirow}
\usepackage{enumitem}
\usepackage{leading}
\usepackage{textcomp}
\usepackage{collcell}
\usepackage{hhline}
\usepackage{pgf}
\newcommand\gray{gray}

\newcommand\ColCell[1]{%
	\pgfmathparse{#1<.8?1:0}%
	\ifnum\pgfmathresult=0\relax\color{white}\fi
	\pgfmathparse{1-#1}%
	\expandafter\cellcolor\expandafter[%
	\expandafter\gray\expandafter]\expandafter{\pgfmathresult}#1}

\newcolumntype{E}{>{\collectcell\ColCell}c<{\endcollectcell}}

\newcommand{\RN}[1]{%
	\textup{\uppercase\expandafter{\romannumeral#1}}%
}

\newcommand{\SystemName}{CapSense\xspace}

\setcopyright{acmlicensed}
\acmDOI{0000001.0000001}

\begin{document}
% Title portion. Note the short title for running heads
\title{Capacitor Based Activity Sensing for Kinetic Powered Wearable IoTs}

\author{Guohao Lan}
\affiliation{%
  \institution{University of New South Wales}
  \city{Sydney}
  \state{NSW}
  \country{Australia}}
\email{guohao.lan@unsw.edu.au}

\author{Dong Ma}
%\authornote{This is the corresponding author}
\affiliation{%
	\institution{University of New South Wales}
	\city{Sydney}
	\state{NSW}
	\country{Australia}}
\email{dong.ma1@student.unsw.edu.au}

\author{Weitao Xu}
\affiliation{%
	\institution{Shenzhen University}
	\city{Shenzhen}
	\country{China}}
\email{weitao.xu@szu.edu.cn}

\author{Mahbub Hassan}
\affiliation{%
	\institution{University of New South Wales}
	\city{Sydney}
	\state{NSW}
	\country{Australia}}
\email{mahbub.hassan@unsw.edu.au}

\author{Wen Hu}
\affiliation{%
	\institution{University of New South Wales}
	\city{Sydney}
	\state{NSW}
	\country{Australia}}
\email{wen.hu@unsw.edu.au}

\begin{abstract}
We propose a novel use of the conventional energy storage component, i.e., capacitor, in kinetic-powered wearable IoTs as a sensor to detect human activities. Since different activities accumulate energies in the capacitor at different rates, these activities can be detected directly by observing the charging rate of the capacitor.  The key advantage of the proposed capacitor based activity sensing mechanism, called CapSense, is that it obviates the need for sampling the motion signal during the activity detection period thus significantly saving power consumption of the wearable device.  A challenge we face is that capacitors are inherently non-linear energy accumulators, which, even for the same activity, leads to significant variations in charging rates at different times depending on the current charge level of the capacitor. We solve this problem by jointly configuring the parameters of the capacitor and the associated energy harvesting circuits, which allows us to operate on charging cycles that are approximately linear. We design and implement a kinetic-powered shoe sole and conduct experiments  with 10 subjects. Our results show that CapSense can classify five different daily activities with 95\% accuracy while consuming 73\% less system power compared to conventional motion signal based activity detection.
\end{abstract}






\maketitle

\renewcommand{\shortauthors}{G. Lan et al.}

\section{Introduction}

The rapid development of embedded technology has enabled wearable IoTs~\cite{seneviratne2017survey} that provide autonomous health and fitness monitoring services, such as step-counting~\cite{consolvo2008activity} and recognition of daily activities~\cite{keally2011pbn}. Such activity detection is achieved by sampling a time series of the motion signal, e.g., the 3-axial accelerations 25-100 times per second ~\cite{bulling2014tutorial,lara2013} depending on the activity detection requirements. As 24/7 health and fitness monitoring becomes essential, high power consumption due to continuous sampling of the motion samples limits the battery life of these wearable devices.

In general, the power consumption in sampling is directly proportional to the sampling rate, as the higher the sampling rate, the more power is consumed by the sensors as well as the microcontroller (MCU), which has to wake up more frequently to read, process, and store the samples. A large volume of past research on context sensing, therefore, has focused on reducing the sampling rates of accelerometer-based systems~\cite{yan2012energy,qi2013adasense}. More recently, researchers have investigated the use of kinetic energy harvesting transducer as a sensor to detect different contexts~\cite{kalantarian2015monitoring,blank2016ball,khalifa2017harke,xuTMCGait,mahbub2018computer}. The instantaneous electric voltage signal generated by the energy harvesting transducer is used as an alternative to the acceleration signal provided by a conventional accelerometer. Transducer-based context-sensing method introduces new power saving opportunities for power-limited wearable devices as, unlike the accelerometers, transducers themselves do not consume any external power. Thus, by saving the energy that would have otherwise consumed by the accelerometer, transducer-based systems can further reduce the sampling power consumption~\cite{khalifa2017harke}. However, as transducer-based approach relies on a time series of signal as the input for activity recognition, it still requires the MCU to frequently wake up and consumes a considerable amount of limited power in energy limited wearable devices.  

In this paper, we propose a new way to detect activities for kinetic energy harvesting powered wearable IoTs, which obviates the need for frequent motion signal sampling and allows very aggressive duty cycling of the MCU to reduce power consumption of wearable devices by several orders. To avoid motion signal sampling, the proposed system, which we call \SystemName, capitalizes two important observations:  
\setlength{\leftmargini}{2em}
\begin{itemize}
	\item[(1)] 	\textbf{The kinetic power of human activities are distinct.} It has been widely demonstrated in the literature that the kinetic energy harvested from different activities are distinctively different~\cite{gorlatova2014movers,yun2008quantitative}. Thus, the energy generation rate of the kinetic-powered wearable device can be used as a feature for human activity recognition.
	\vspace{0.1in}
	\item[(2)] \textbf{Capacitor provides accumulated information}. In kinetic powered devices, the energy generated by the energy harvesters are naturally stored in the associated capacitor. More importantly, the capacitor charging rate provides information about the \textit{energy generation rate} of the external activity. Interestingly, charging rate of the capacitor can be obtained by simply reading the capacitor voltage at the end of each activity detection period, which is typically about 5 seconds~\cite{lara2013survey,bulling2014tutorial}, without the need of sampling the instantaneous signal many times during this period.  	
\end{itemize}
Thus, it should be possible to classify human activities by simply reading the capacitor voltage only once in every 5 seconds. Comparing with conventional motion signal based activity detection, which requires the system to wake up many times per second~\cite{lara2013}, CapSense allows very aggressive duty cycling of the embedded IoT. However, in realizing CapSense, we face two challenges. The first challenge we face is that, even for the same activity, CapSense leads to significant variations in charging rates at different times depending on the current charge level of the capacitor. This is because of the fundamental charging property of capacitors, which dictates that it becomes harder to charge a capacitor as it accumulates more charges~\cite{westerlund1994capacitor}. The second challenge arises due to use of a \textit{simple} and \textit{single} variable/feature, i.e., the capacitor charging rate, for classifying all activities in contrast to many motion samples and features used by conventional activity detection. Both challenges must be addressed to realize acceptable activity classification accuracy with CapSense. 

The contributions of this paper can be summarized as follows:
\begin{itemize}
	\item[(1)] We propose a new method for human activity sensing, \SystemName, which detects activity from the charging rate of the energy storing capacitor. To the best of our knowledge, such capacitor-based activity detection has not been explored before.
	\item [(2)] We address the first challenge of non-linear capacitor charging by jointly configuring the parameters of the capacitor and the associated energy harvesting circuits, which allows us to operate with capacitor charging cycles that are approximately linear. 
	\item[(3)] We implement the idea of \SystemName in shoe form factor using piezoelectric bending energy harvester. We address the single feature classification challenge by introducing two energy harvesters and capacitors, one at the rear of the sole and the other at the front. We show that the proposed dual-capacitor system significantly improves classification performance of CapSense as it can effectively differentiate activities by leveraging the energy generation difference between the rear and front of the foot.
	
	\item[(4)] Using our dual-capacitor prototype, we conducted experiments with 10 subjects performing 5 different activities. We demonstrate that \SystemName is capable of detecting daily activity with up to 95\% accuracy.
	\item[(5)] We conduct a detailed power profiling to quantify the power saving opportunity of \SystemName. Our measurement results indicate that, compared to the state-of-the-art, \SystemName reduces sampling-related power consumption by 54\% and the overall IoT system power consumption by 73\%.
\end{itemize}

Partial and preliminary results of this paper have appeared in our previous work~\cite{lan2017capsense}. In this paper we provide the following two major extensions to the conference version: (1) We redesign the previous \SystemName prototype by adding a second energy harvester and capacitor to the shoe insole (at the front) thus realizing the proposed \textit{dual-capacitor} wearable prototype, and (2) We conduct new sets of experiments and collect a new dataset for the dual-capacitor prototype. Using the new dataset, we demonstrate that fusing data from two capacitors improves activity recognition accuracy by up to 11\% compared to single capacitor systems. 

The rest of the paper is organized as follows. We first introduce some background of kinetic-powered IoT in Section~\ref{section:background}. Then, we present the design and implementation of \SystemName in Section~\ref{section:system_overview}, followed by its performance evaluation in Section~\ref{section:system_evaluation_keh}. The power measurement study is presented in Section~\ref{section:energy_consumption}. We review the related works in Section~\ref{section:related_work} before we conclude our work in Section~\ref{section:conclusion}.

\section{Preliminaries in kinetic powered IoT}
\label{section:background}

In this section, we provide some basic background of kinetic-powered IoTs and the concept of using kinetic energy harvesting transducer for sensing.

\subsection{Kinetic Energy Harvesting}

\begin{figure}[]
	\centering	\includegraphics[scale=0.85]{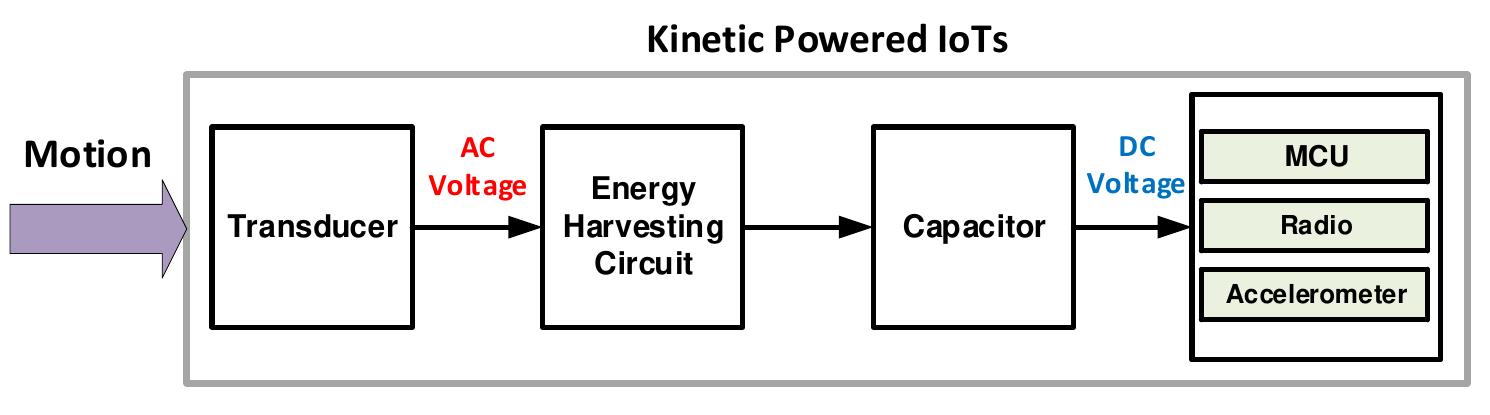}
	\caption{Generic architecture for kinetic powered IoT device.} \label{Fig:KEH_Architect}
	\vspace{-0.2in}
\end{figure}

Kinetic energy is the energy of an object due to its motion. Kinetic energy harvesting refers to the process of scavenging kinetic energy released from human activity or ambient vibrations. The use of kinetic energy harvesting for self-powered IoT has been widely investigated in the literature~\cite{mitcheson2008energy,bhatti2016energy}. Figure~\ref{Fig:KEH_Architect} shows a generic architecture for a kinetic powered IoT device which typically contains a transducer, i.e., energy generator, that can convert mechanical energy into electric AC voltage, and a set of energy harvesting circuit that converts the AC voltage into regulated DC output, and a energy storage element, e.g., a capacitor, to store the harvested energy. The stored energy will be used to power external user loads (e.g., sensors, MCU, or radio) when sufficient amount of energy has been accumulated. 

\begin{figure}[]
	\centering
	\subfigure[Inertial-force transducer]{
		\includegraphics[scale=0.75]{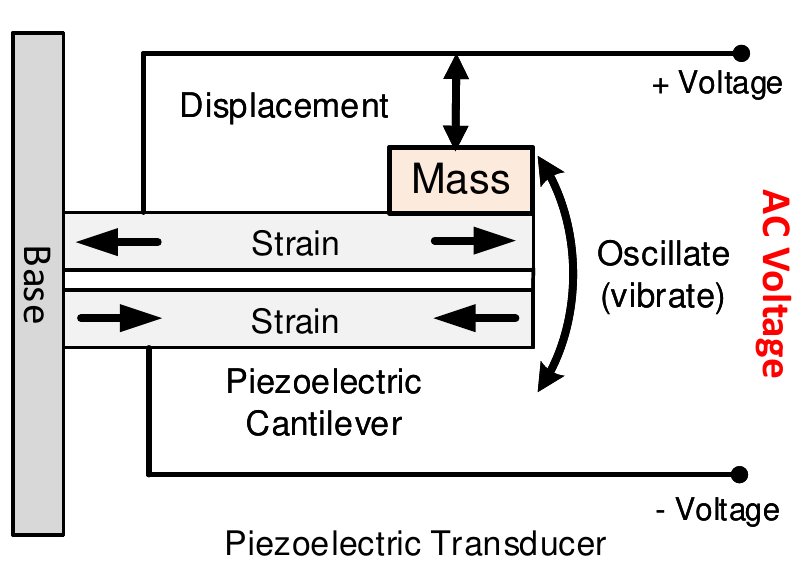}
		\label{fig:inertial_base}}
	\subfigure[Direct-force transducer.]{
		\includegraphics[width=2.5in]{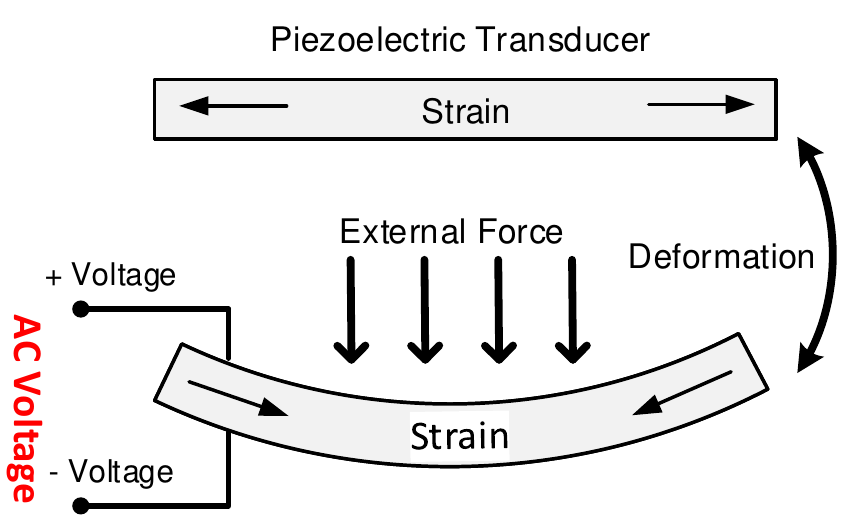}
		\label{fig:force_base}
	}
	\caption{Principle of kinetic energy harvesting transducer.}
	\label{Fig:transducer_principle}
	\vspace{-0.15in}
\end{figure}

There are three main energy transduction techniques that are widely used in the literature, namely, \textit{piezoelectric}, \textit{electromagnetic}, and \textit{electrostatic}. Among them, piezoelectric is the most favourable transduction mechanism for wearable IoTs, due to its simplicity and compatibility with MEMS (micro electrical mechanical system). But, fundamentally, the three techniques share the same physical mechanism to covert kinetic energy into electric power. Depending on the energy harvesting scenario, transducer can be classified into two different categories: the inertial-force transducer and direct-force transducer~\cite{mitcheson2008energy}. As shown in Figure~\ref{Fig:transducer_principle}(a), the inertial-force transducer is usually modeled as an inertial oscillating system consisting of a cantilever beam attached with two piezoelectric outer-layers. One end of the beam is fixed to the device, while the other is set free to oscillate (vibrate). When the piezoelectric cantilever is subjected to a mechanical stress, it expands on one side and contracts on the other. The induced piezoelectric effect will generate an AC voltage output as the beam oscillates around its neutral position. Similar, in terms of the direct-force transducer shown in Figure~\ref{fig:force_base}, AC voltage signal is generated when the piezoelectric transducer is deformed (bended) due to the external mechanical force. In this paper, we build our proof-of-concept prototype based on the direct-force based piezoelectric transducer (in Section~\ref{sec:prototype_design}). %We use the AC voltage generated by the piezoelectric transducer as the signal source for transportation mode detection.

\subsection{KEH Transducer-based Sensing}

Although KEH transducer is designed with the purpose of scavenging kinetic energy from motions, researchers have investigated the use of KEH transducer as a low power vibration sensor for context detection~\cite{li2013powering,kalantarian2015monitoring,khalifa2017harke}, in which, the AC voltage generated by the transducer is used as the signal for sensing. Comparing with conventional vibration sensor, e.g., accelerometer, KEH transducer-based system is able to eliminate the energy consumed in powering the accelerometer. For instance, in~\cite{khalifa2017harke}, a KEH-transducer based activity recognition system is designed. The proposed system is able to achieve $83\%$ of accuracy for classifying different daily activities while saving 79\% of power that will be consumed by an accelerometer. 

%In a similar work proposed by Kalantarian et al. ~\cite{kalantarian2015monitoring}, a piezoelectric transducer-based wearable necklace was utilized to monitor the eating habits of the user

\section{System Overview}
\label{section:system_overview}
In this section, we present the concept, design, and implementation of \SystemName.%, an ultra-low power daily activity sensing system for self-powered wearable devices.

\subsection{\SystemName Concept}

\begin{figure}[]
	\centering	\includegraphics[scale=0.83]{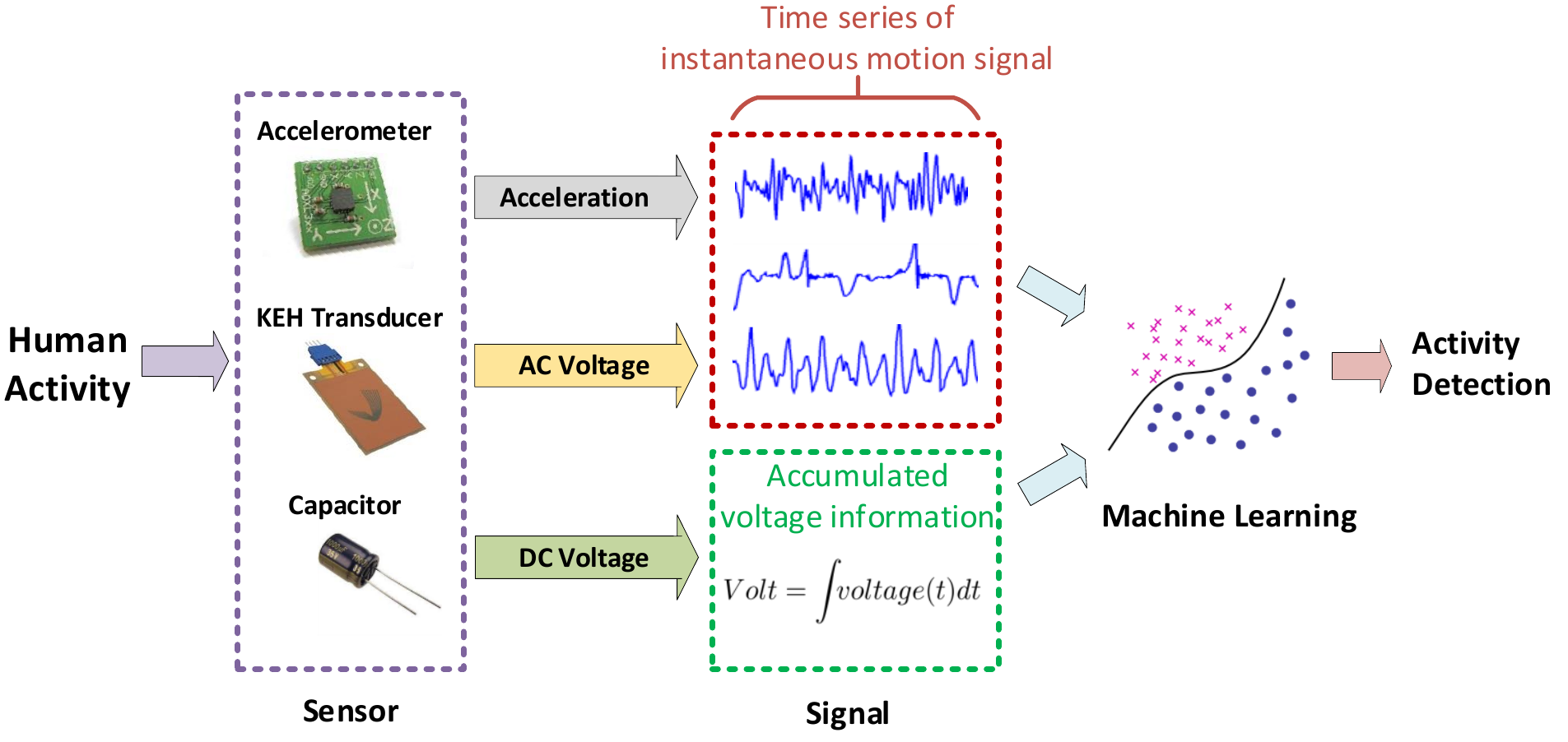}
	\caption{The processing pipeline of a typical activity sensing system. Comparing with both accelerometer and KEH-transducer-based system, \SystemName does not require any time series of motion signal for classification.} \label{Fig:idea_compare}
	\vspace{-0.1in}
\end{figure}

%\subsubsection{The dilemma of energy consumption in signal acquisition}
Figure~\ref{Fig:idea_compare} exhibits the processing pipeline of a typical activity sensing system. It usually consists a sequence of procedures, including the acquisition of motion signal from sensor, signal processing, feature extraction, and utilizing machine learning algorithms for classification. We can notice that for both accelerometer and KEH-transducer based sensing systems, \textit{\textbf{a time series of instantaneous motion signal}} (either acceleration or AC voltage signal) during a given activity detection period is required as the input for classification. This time series of motion signal is usually obtained by sampling the motion sensor at a frequency of 25-100Hz depending on the classification accuracy required~\cite{bulling2014tutorial}. This implies that, the MCU of the IoT device must wake up at least 25-100 times per second to acquire the instantaneous motion signal. As we will demonstrate later in the paper (see Section~\ref{section:energy_consumption}), a large fraction of the sampling power is actually consumed due to waking up the MCU. Although the use of KEH-transducer can eliminate the energy consumption in powering accelerometers, it still needs to continuously sense and process the AC voltage signal from the KEH transducer at a high sampling rate, and thus, it continues to face the energy consumption problem and consume a significant amount of the harvested energy in kinetic powered devices. This motivates the design of \SystemName, which aims to achieve high accuracy human activity sensing while eliminating the need of time series of motion signal.

As shown in Figure~\ref{Fig:idea_compare}, unlike accelerometer or KEH transducer-based systems that require \textit{a time-series of instantaneous motion signal} sampled from the sensor at a high frequency, \SystemName utilizes \textit{a single sample} of the capacitor voltage for activity recognition. The feasibility of \SystemName relies on two fundamental facts: 
\setlength{\leftmargini}{4.5em}
\begin{itemize}
	\item[Fact 1.] \textbf{The kinetic power of human activities are distinct}. It has been widely demonstrated in the literature that the kinetic power harvested from different activities are different~\cite{gorlatova2014movers,yun2008quantitative}. Thus, the energy generation rate of the kinetic-powered wearable device can be used as a feature for human activity recognition.
	\vspace{0.1in}
	\item[Fact 2.] \textbf{Capacitor provides accumulated information}. As shown previously in Figure~\ref{Fig:KEH_Architect}, in kinetic powered devices, the energy generated by KEH transducer are naturally stored in the associated capacitor. More importantly, the capacitor charging rate provides information about the \textit{energy generation rate} of the external activity. Interestingly, as it will be shown in Section~\ref{sec:activity_sensing_capacitor_voltage}, charging rate of the capacitor over the last $T_C$ second(s) can be estimated by simply reading the capacitor voltage once every $T_C$ second(s), without the need of sampling the AC voltage signal frequently to calculate the average harvesting power. 	 
\end{itemize}

Those two facts imply that by leveraging the capacitor voltage change over a time period of $T_C$, we can estimated the corresponding energy generation rate and leverage it to recognize the activity performed by the user in the last $T_C$ period of time. As we will demonstrate later in Section~\ref{section:system_evaluation_keh}, \SystemName can detect activities by reading the capacitor voltage once every $T_C$=5 seconds compared to tens of Hz required by the state-of-the-art~\cite{bulling2014tutorial,khalifa2017harke}. The fundamental novelty of \SystemName is that, unlike accelerometer or KEH transducer that can only generate instantaneous motion information of the subject at a particular time, capacitor accumulates the generated KEH energy over time, and the capacitor voltage provides \textbf{\textit{accumulated information}} of the subject over a period of time. Thus, it allows very aggressive duty cycling of the MCU and reduces sensing-induced power consumption of wearable devices by several orders. In the following, we introduce the design and implementation of \SystemName.

\subsection{\SystemName Architecture}

\begin{figure}[]
	\centering	\includegraphics[scale=0.8]{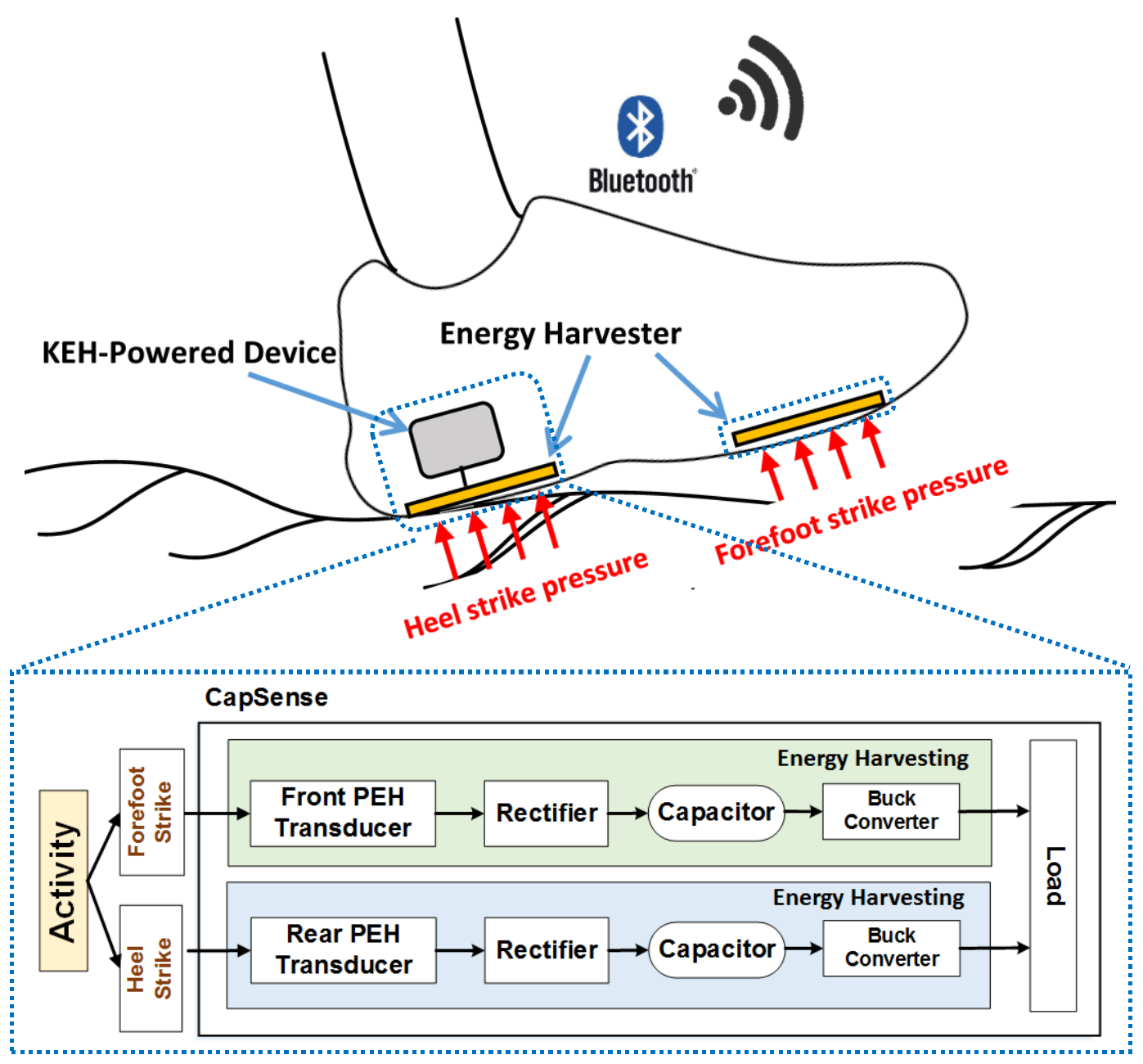}
	\caption{System architecture of \SystemName.} \label{Fig_system_overview}
	\vspace{-0.1in}
\end{figure}

There are a various of design options in kinetic energy harvesting powered wearable devices, such as backpack~\cite{xie2014human}, fabric~\cite{yang2012piezoelectric}, wristband~\cite{SeikoKinetic}, and footwear~\cite{kymissis1998parasitic,shenck2001energy}. We designed our system in the form-factor of shoes for several reasons: first, shoes are worn by users for the majority of time in their daily lives; second, unlike many other wearable devices, shoes have a much larger space to place the energy harvesters; third, is known that feet can generate more energy than other body parts, such as wrist~\cite{ishida2013insole,antaki1995gait}. 

Figure~\ref{Fig_system_overview} exhibits the system architecture of \SystemName and  visualizes how it works. Considering the scenario in which a subject is wearing the KEH-powered shoes and doing some activities, e.g., walking or running, her foot will hit the ground floor and the pressure induced by both the heel and forefoot strikes will bend the two piezoelectric energy harvesters (PEH) inside the shoe accordingly. Consequently, the energy harvesters generate electric power from the foot strikes when the subject is doing different activities, and the generated energy are naturally accumulated in the associated capacitor. As discussed, since the power generated from different human activities such as walking, running, and relaxing, are distinctively different~\cite{gorlatova2014movers}, and the energy generated by the PEH transducers within the last $T_C$ second(s) is accumulated in the capacitors, it would be possible to estimate the power generation rate during $T_C$ by a single sample of the capacitor voltage. Then, the estimated rates can be used to recognize the activity performed by the user in the last $T_C$ second.

As shown in Figure~\ref{Fig_system_overview}, \SystemName consists of two parts: \textit{Load} and~\textit{Energy Harvesting}. \textit{Load} represents any system components responsible for data sensing, processing, and communication, or could be a rechargeable battery that can be used to power a wearable system. The \textit{Energy Harvesting} corresponds to the functional components that harvest and accumulate energy from human activity. It includes two piezoelectric energy harvesting (PEH) transducers, i.e., front and rear PEH transducers, to harvest energy from the foot strikes. As exhibited in Figure~\ref{Fig_system_overview}, the front PEH transducer is able to harvest energy from the ground reaction pressure associated with the toe-off phases when the forefoot hit the ground, whereas, the rear PEH transducer is designed for harvesting energy from the pressure caused by the heel strikes. More specifically, when the subject is walking/running, the body weight induced pressure will be transferred to the PEH transducers through bones~\cite{kong2008smooth}. Given the anatomical structure of the foot, the front PEH is placed at the location to capture the pressure transfered from the Metatarsal bones, while the rear PEH is placed to capture the pressure from the Calcaneus (Heel) bone. In \SystemName, the two PEH transducers are then connected to the rectifying circuit to rectify the intermittent or continuous AC voltage output from the PEH transducers into stable DC power. The rectified DC voltage will be accumulated in the corresponding capacitor before it is sufficient to turn-on the \textit{buck converter} and power any electronics (i.e., the voltage of the capacitor should reach a pre-defined threshold configured in the buck converter). As it will be discussed in Section~\ref{section:system_evaluation_keh}, by fusing the voltage signal of the two capacitors, we can significantly improve the sensing accuracy.

\subsection{\SystemName Prototype Design}
\label{sec:prototype_design}

\begin{figure}[]
	\centering
	\subfigure[\SystemName prototype.]{
		\includegraphics[scale=0.6]{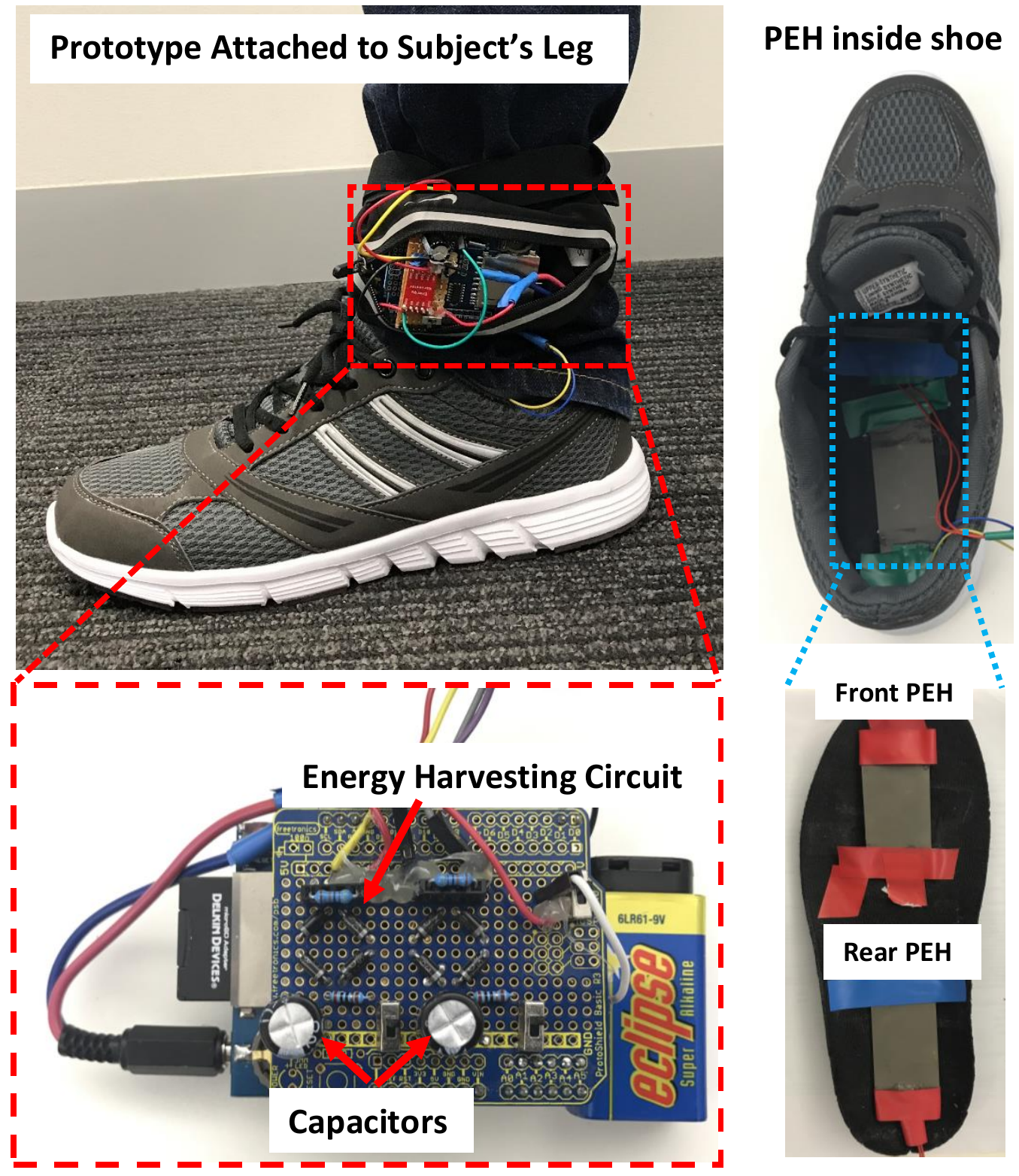}
		\label{Fig:PEHhardware}}
	\subfigure[PEH Transducers.]{
		\includegraphics[scale=0.45]{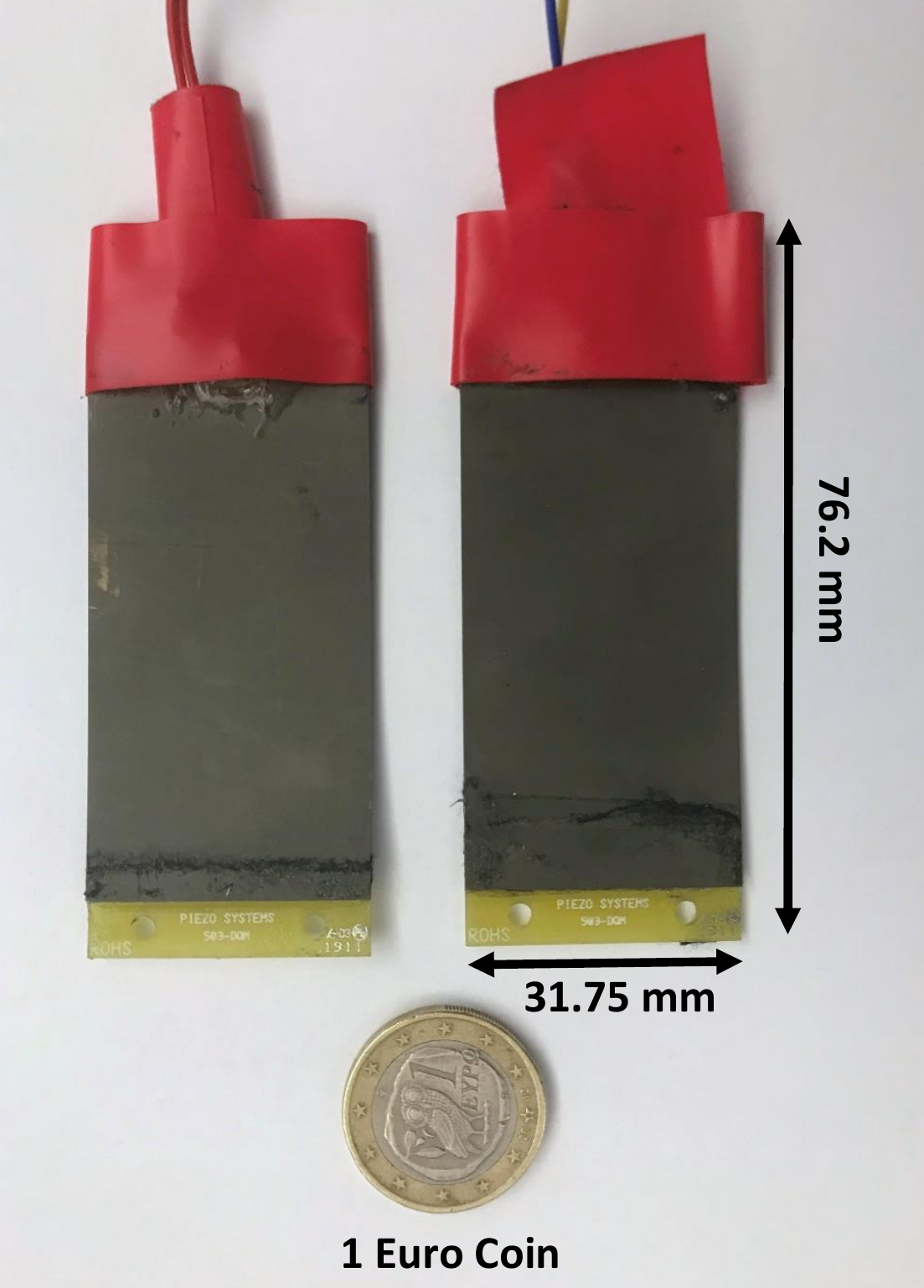}
		\label{Fig:PEHS}}
%	\subfigure[overall circuit design.]{
%		\includegraphics[scale=0.72]{Figures/circuit_diagram.pdf}
%		\label{Fig:DataCollectionCircuit}}
	\caption{\SystemName prototype design.}
	\vspace{-0.1in}
\end{figure}

In the following, we present the design and implementation of \SystemName. Figure~\ref{Fig:PEHhardware} gives the pictures of our prototype which we implemented in the form of shoe. As discussed previously, our prototype consists of two parts, the \textit{Energy Harvesting} and \textit{Load}. For the Energy Harvesting part, we select two EH220-A4-503YB PEH bending transducers from Piezo Systems\footnote{Piezo System: http://www.piezo.com/prodexg8dqm.html.} as our PEH transducers and attached them to the shoe-pad. The transducers are only 10.4 grams each with a dimension of 76.2$\times$31.75$\times$2.28 $mm^3$, which makes it easy to be placed inside the shoe. As shown in Figure~\ref{Fig:PEHhardware}, the two PEH transducers are fixed at the front and rear of the shoe-pad to harvest energy from the heel and forefoot strikes, respectively. The details of the PEH transducers are given in Figure~\ref{Fig:PEHS}. 

The output pins of the PEH transducer are connected to an energy harvesting circuit, namely the LTC3588-1 from the Linear Technology\footnote{LTC3588: http://www.linear.com/product/LTC3588-1.}. The LTC3588-1 integrates a low power-loss bridge rectifier that can be used to rectify the AC voltage output from the PEH transducer, and a high efficiency buck converter that is able to transfer the energy stored in the capacitor into stable DC power to power/charge the load. We select two electrolytic capacitors with a capacitance of 470$\mu$F and a rating voltage (i.e., the maximum voltage) of 25V to store the generated energy from the two PEH transducers (we will discuss how we select the capacitor in Section~\ref{sec:properties_capacitor}). When the voltage of the capacitor rises above the undervoltage lockout rising threshold of the buck converter (i.e., denoted by $V_{UVLO}$ $rising$, and equals to $4V$ in our setting), the buck converter will be enabled to discharge the energy stored in the capacitor. On the other hand, when the capacitor voltage has been discharged below the lockout falling threshold (i.e., denoted by $V_{UVLO}$ $falling$, and equals to 3.08$V$ in our setting), the buck converter will be turned off, and the capacitor starts to accumulate any harvested energy. 

\begin{figure}[]
	\centering
	\includegraphics[scale=0.8]{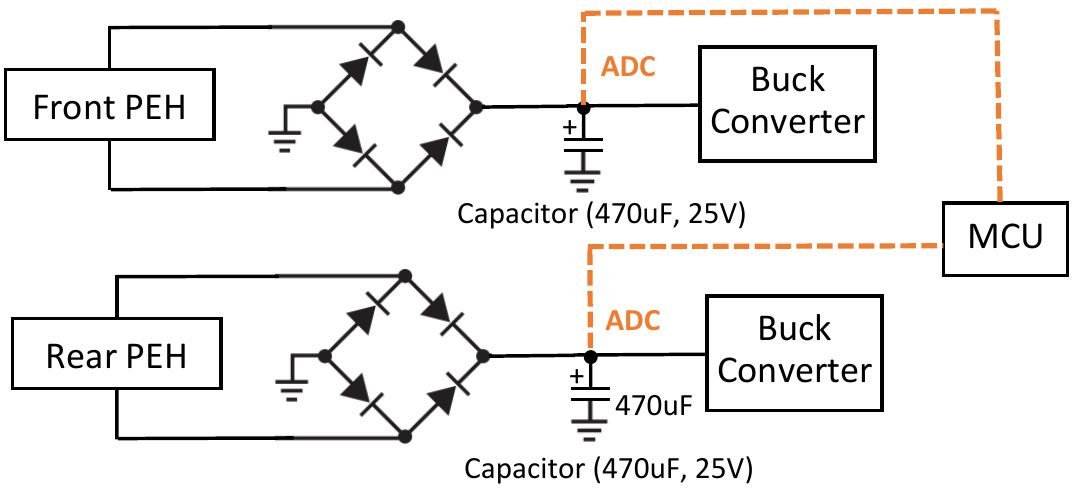}
	\caption{\SystemName overall circuit design.}
	\label{Fig:DataCollectionCircuit}
	\vspace{-0.1in}
\end{figure}

The simplified circuit diagram of the prototype is shown in Figure~\ref{Fig:DataCollectionCircuit}. The front and rear capacitors are used to store the energy harvested from the front and rear PEH, respectively. For analysis purpose, we use an Arduino Uno board to sample the voltage of the two capacitors through the onboard 10-bit ADC at 100 Hz and stored the sampled data on the SD card for offline data analysis. By leveraging this data, we will analyze the performance of \SystemName, and demonstrate that it is able to achieve over 90\% classification accuracy by sampling the capacitor voltage once every five seconds.

\subsection{Ensuring Linearity in Capacitor Voltage}
\label{sec:properties_capacitor}

Before presenting the details of capacitor-based sensing, we first analyze some properties of the capacitor when the system is powered by an energy harvester, and discuss the feasibility and design requirement of leveraging the capacitor voltage for activity sensing. 

The voltage of the capacitor, $V_{C}(t)$, at time $t$ during the charging is given by:
\begin{equation}
V_{C}(t)=V_{max}(1-e^{-t/\tau}),
\label{E1}
\end{equation}
in which, $V_{max}$ is the maximum voltage to which the capacitor can be charged, and it is bounded by $V_{max} = min\{V_{rating}, V_S\}$, where $V_{S}$ is the voltage applied to the capacitor, i.e., the rectified DC voltage from the rectifier in our case, and $V_{rating}$ is the rating voltage of the capacitor; and $\tau$ is defined as $\tau = RC$, in which, $R$ is the resistance of the resistor in the equivalent resistor-capacitor charging circuit (RC circuit), and $C$ is the capacitance of the capacitor. For a given capacitor, $\tau$ is known as the \textit{time constant} of the equivalent RC circuit, which is a constant value (in second).

\begin{figure}
	\centering
	\includegraphics[scale=0.62]{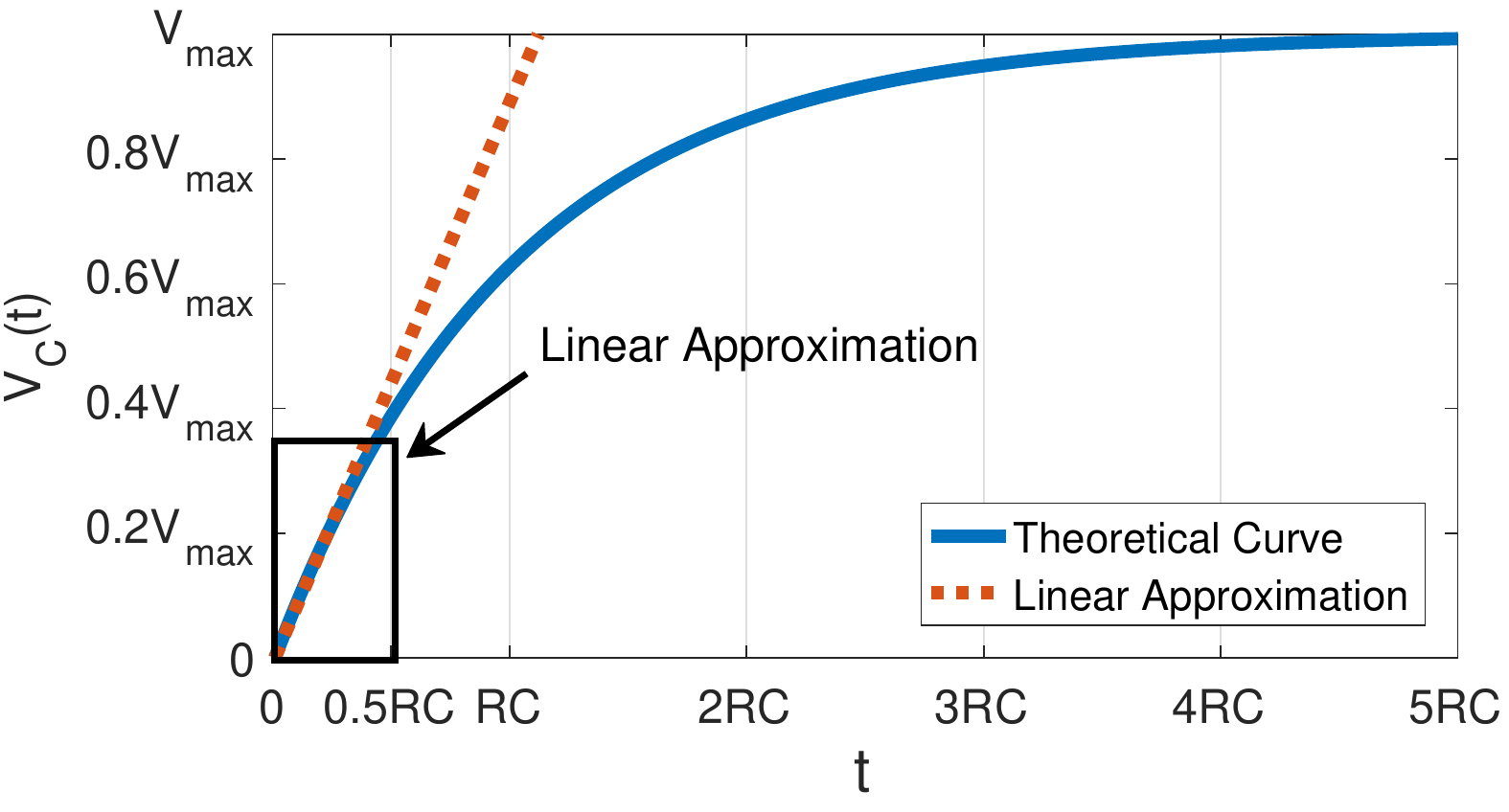}
	\caption{The theoretical curve exhibits the voltage of the capacitor $V_{C}(t)$ when it is charged over time $t$. The theoretical curve is approximated by an linear curve. The unit of $t$ is $RC$ (i.e., the time constant). } 
	\label{fig:linear_approximation}
	\vspace{-0.1in}
\end{figure}

The relation between the capacitor voltage, $V_{C}(t)$, and time $t$ is visualized in Figure~\ref{fig:linear_approximation}. The theoretical curve indicates the voltage of the capacitor when it is charged by the supply power $V_S$ over time (in our case, $V_S$ is the rectified DC voltage from the rectifier). The first observation is that, within the examining time of 5$RC$, $V_{C}(t)$ increases exponentially and the increasing rate of capacitor voltage is not constant. For instance, the voltage increment in the first $RC$ interval (i.e., from time 0 to $RC$) is not equal to that increased in the second $RC$ interval (i.e., from time $RC$ to $2RC$). As we will discuss later in Section~\ref{sec:activity_sensing_capacitor_voltage}, the only information we can obtain from the capacitor is its voltage and we are using the voltage increasing rate for activity sensing, the nonlinear increment in the capacitor voltage will introduce additional uncertainties in the voltage increasing rate, and thus, impairs the activity recognition accuracy. We should therefore carefully select a suitable capacitor to ensure the linearity of the capacitor voltage during our sensing. 

Fortunately, as we can observe in Figure~\ref{fig:linear_approximation}, with time $t \leq \frac{1}{2}RC$, the theoretical curve of $V_{C}$ (defined in Equation~\ref{E1}) can be approximated by a linear curve~\cite{macdonald1955charging}. According to the RC circuit theory, a capacitor can be charged to 39.3\% of $V_{max}$ with a charging time of $\frac{1}{2}RC$, this means that to ensure the linearity in the capacitor voltage for a maximum time of $\frac{1}{2}RC$, $V_{max}$ should satisfy:
\begin{equation}
%\dfrac{V_{UVLO}}{V_{max}} \leq 0.393,
V_{max} \geq \dfrac{V_{UVLO}\ rising}{0.393},
\label{E2}
\end{equation}
which yields:
\begin{equation}
%\dfrac{V_{UVLO}}{V_{max}} \leq 0.393,
min\{V_{rating}, V_S\} \geq \dfrac{V_{UVLO}\ rising}{0.393},
\end{equation}
in which, $V_{UVLO}\ rising$ is the undervoltage lockout rising threshold of the buck converter, at which the capacitor starts to be discharged. This means that, to ensure the linearity in the capacitor voltage, the selection and configuration of the hardware components (i.e., the PEH transducer, the buck converter, and capacitor) should be considered interactively. 

As discussed in Section~\ref{sec:prototype_design}, in our prototype, $V_{UVLO}\ rising$ of the buck converter has been configured to 4V\footnote{According to the datasheet of LTC3588, to ensure an output DC voltage of 2.5V, the lowest voltage threshold is 4V.}. This means that, given Equation (3), we have: $min\{V_{rating}, V_S\} \geq \frac{4 V}{0.393}=10.18V$. The rectified DC voltage from the rectifier, $V_S$, depends on the energy harvester that is used in the system. Given different materials and configurations of the energy harvester, $V_S$ could be as high as tens of volts. In our case, the rectified voltage from the PEH transducer we used in the current prototype is up to 20.8V which is much higher than 10.18V. Therefore, it turns out that the rating voltage of the capacitor, $V_{rating}$, should be larger than 10.18V. We select a capacitor with rating voltage of 25V to meet the requirement. 

\subsection{Activity Sensing using Capacitor Voltage}
\label{sec:activity_sensing_capacitor_voltage}

\begin{figure}[]
	\centering
	\includegraphics[scale=1]{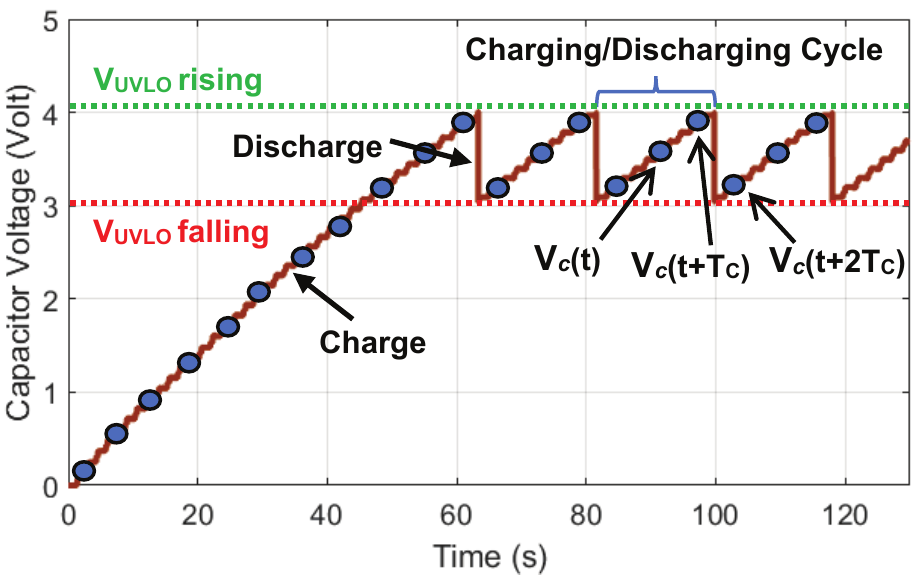}
	\caption{Voltage trace showing the capacitor is charged and discharged periodically. The blue dot dots indicate possible sampling point at which the MCU wakes up to read the capacitor voltage.} 
	\label{fig:capacitor_charge} 
	\vspace{-0.1in}
\end{figure}

In the following, we discuss how to leverage the capacitor voltage for activity sensing. Figure~\ref{fig:capacitor_charge} exhibits an actual voltage trace showing the charging and discharging cycles of the capacitor when it is powered by the energy harvester. The charging/discharging behavior of the capacitor is controlled by the energy harvesting circuit depending on the capacitor voltage level. Initially, the capacitor voltage starts from 0, and takes approximately 60 seconds to reach 4V and triggers the buck converter to discharge the accumulated energy (i.e., the $V_{UVLO}$ of the buck converter is 4V). Then, when the capacitor voltage is discharged to 3.08V, the buck converter is shut off until the capacitor voltage reaches the $V_{UVLO}$ rising threshold again.

During the charging/discharging of capacitor, \SystemName duty-cycles the system MCU to periodically sample the capacitor voltage. As an example shown in Figure~\ref{fig:capacitor_charge}, MCU wakes up to sample the capacitor voltage once every $T_C$ seconds. Using any two adjacent voltage samples, it is straightforward to estimate the capacitor voltage increment rate, $r$, over the last accumulation time of $T_C$, by: 
\begin{equation}
r = \frac{V_C(t+T_C)-V_C(t)}{T_C}, ~s.t.~~V_C(t+T_C) > V_C(t)
\label{E4}
\end{equation}
in which, $V_C(t)$ and $V_C(t+T_C)$ is the capacitor voltage at time $t$ and $t+T_C$, respectively. 
Therefore, by periodically sampling the capacitor voltage at a frequency of $\frac{1}{T_C}$Hz, we can estimate the energy generation rate of the PEH transducer (either the front or the rear PEH transducer in our prototype) over the last $T_C$ seconds. Note that, as MCU has no knowledge about the charging/discharge status of the capacitor, it is possible that those two adjacent voltage samples are obtained in two different charging cycles. For instance, as shown in Figure~\ref{fig:capacitor_charge}, it may result in $V_C(t+T_C) \geq V_C(t+2T_C)$, as $V_C(t+2T_C)$ is sampled at the initial charging state of the capacitor in a new charging cycle. In this case, when calculate the voltage increment rate, $r$, we disregard all adjacent voltage peer $\{V_C(t), V_C(t+T_C)\}$ that $V_C(t+T_C) \leq V_C(t)$.

\begin{figure}[]
	\centering
	\includegraphics[scale=0.88]{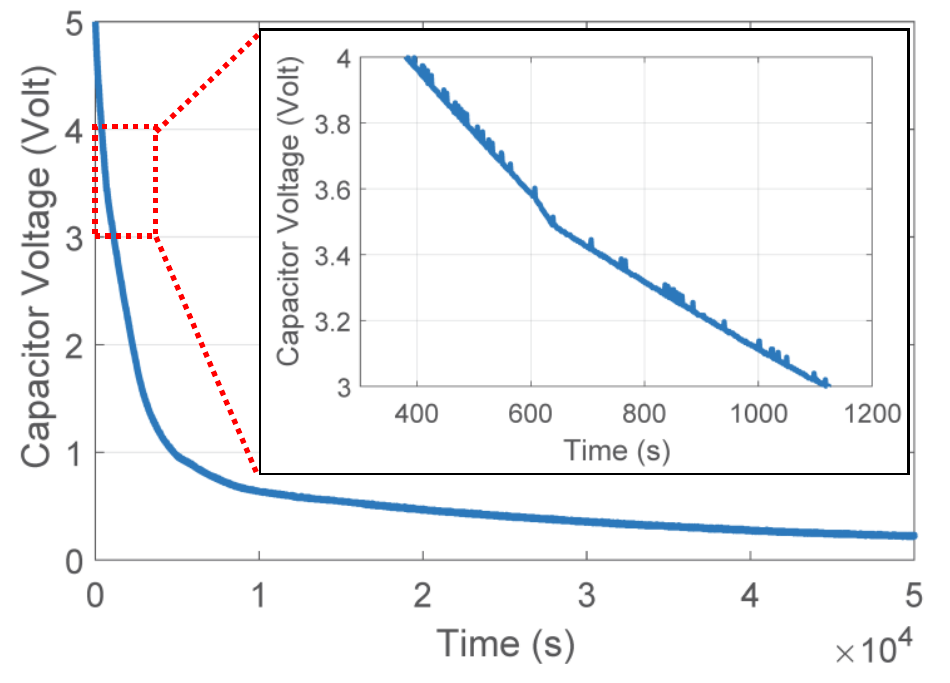}
	\caption{The measured voltage of a self-discharge capacitor.} 
	\label{fig:capacitor_leakage}
	\vspace{-0.1in}
\end{figure}

\subsubsection{Impact from the capacitor discharging}
However, the estimation of $r$ may affect by the discharging time of the capacitor. As if it takes a long time for the buck converter to discharge the capacitor from 4V to 3V, it is possible that the MCU wakes up and samples an incorrect voltage value during the discharge of the capacitor. Which means, the capacitor discharge happens in the middle of the last $T_C$ seconds, such that the actual energy accumulation time is shorter than $T_C$ and part of the accumulated energy has been discharged. Consequently, the voltage increment rate $r$ obtained from Equation~\ref{E4} will be underestimated. Fortunately, as shown in Figure~\ref{fig:capacitor_charge}, the time required by the buck converter to discharge the capacitor from 4V to 3V is less than 10 $m$s according to our measurement. This fast discharging speed ensures that whenever a capacitor discharge happens within the last $T_C$ seconds, the measured voltage $V_C(t+T_C)$ will be much lower than $V_C(t)$, and thus, will be disregard during the estimation.

\subsubsection{Impact from the capacitor self-discharge}
Another factor that may affect the estimation of $r$ is the self-discharge of the capacitor. That is, the voltage leakage of the capacitor when it is not charged by the energy harvester. A high voltage leakage may result in an underestimation in $r$, as part of the harvested energy is lost due to capacitor self-discharge. To investigate the influence of capacitor leakage, we measured the voltage of our capacitor when it self-discharges from 5V to 0V. The results are exhibit in Figure~\ref{fig:capacitor_leakage}. Specifically, we are interested in the behavior of the capacitor within the 3-4V voltage range. As visualized in the amplified subfigure, we can observe that it takes more than 700 seconds (i.e., 12 minutes) for the capacitor to self-discharge from 4V to 3V, which means the leakage of the capacitor is negligible within a short time of a few seconds. Thus, the capacitor leakage will have a very limited impact on our estimation given a few seconds accumulation time $T_C$. 

%\begin{figure}[]
%	\centering
%	\subfigure[The voltage traces of the Front Capacitor which is charged by the Front PEH transducer during activities.]{
%		\includegraphics[scale=0.45]{Figures/charging_front.pdf}
%		\label{fig:charging_front}}
%	\subfigure[The voltage traces of the Back Capacitor which is charged by the Back PEH transducer during activities.]{
%		\includegraphics[scale=0.45]{Figures/charging_back.pdf}
%		\label{fig:charging_back}
%	}
%	\caption{An illustration of the capacitor voltage traces when the subject is doing different activities.}
%	\label{Fig:capacitor_charging_example}
%	%\vspace{-0.1in}
%\end{figure}

%In addition, we can also notice that, the amount of energy that can be harvested from the front PEH is much larger than that from the back PEH. This is mainly because more pressure are transfered to front PEH through the 

\subsubsection{Putting all together}
\begin{figure}[]
	\centering
	\includegraphics[scale=0.68]{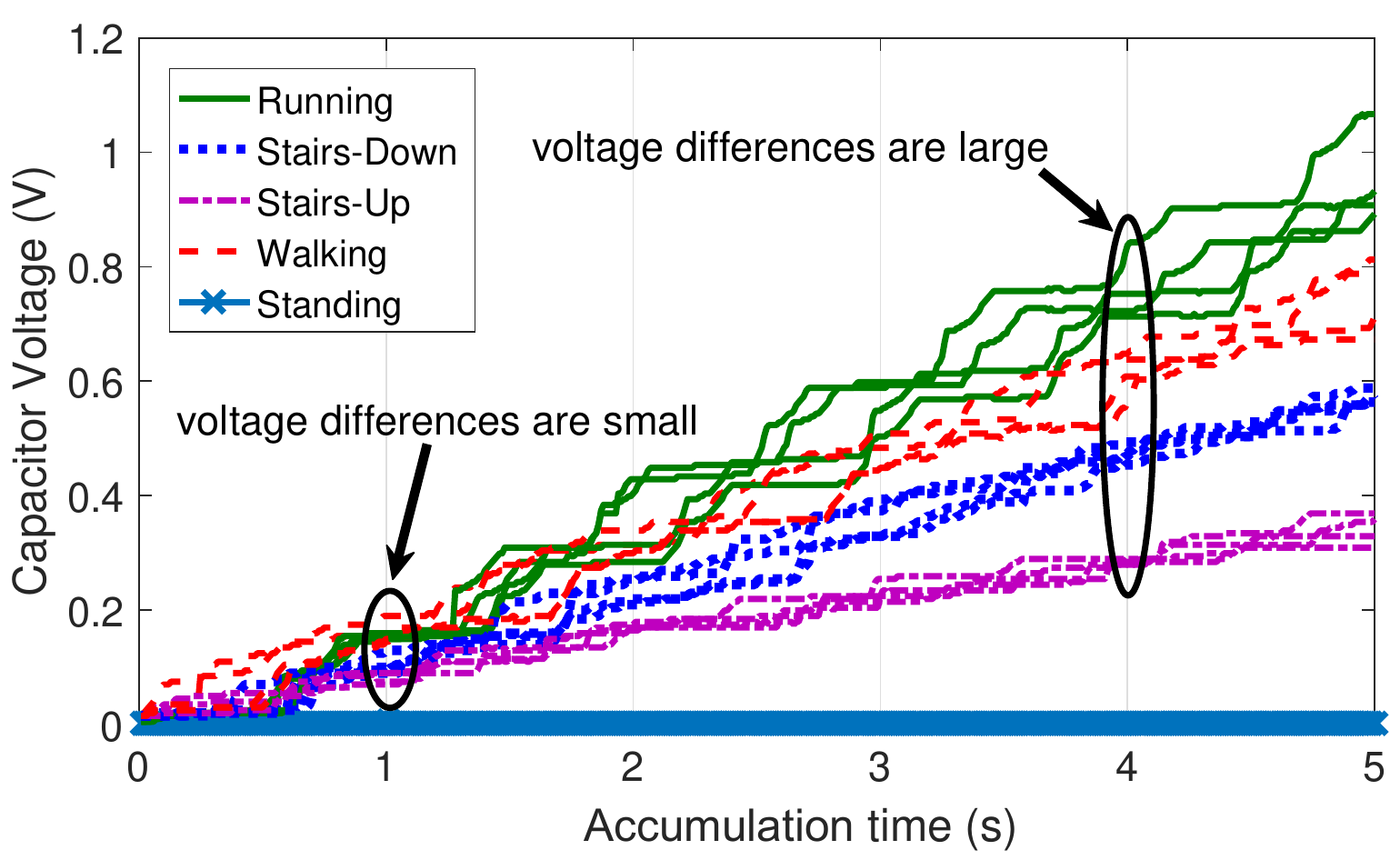}
	\caption{An illustration of the capacitor voltage traces when the subject is doing different activities. The four traces for the same activity are plotted in the same line style. Differences in capacitor voltages for different activities grow with the accumulation time, making it easier to classify activities with larger voltage sampling intervals.} 
	\label{Fig:capacitor_charging_example}
	%\vspace{-0.1in}
\end{figure}

As an example, Figure~\ref{Fig:capacitor_charging_example} plots the voltage traces of the capacitor when a subject is doing different activities. We can observe that, for all five activities, our hardware design ensures an approximately linear increase in the voltage when the capacitor is powered by the energy harvesters. As the energy generation rates of different human activities are distinct, and the increment of capacitor voltage can directly yield the energy generation rate during the last few seconds, thus, we can achieve activity recognition by simply using the capacitor voltage. \SystemName utilizes solely the capacitor voltage to classify different human activities, which enables system level power saving by enabling the MCU to stay in the energy-saving low-power mode for extended periods of time. 

We can also notice from Figure~\ref{Fig:capacitor_charging_example} that the difference in capacitor voltages among different activities increases with the accumulation time. For instance, with $T_C$=1s, the voltage increment rates are similar among different activities. Since \SystemName merely uses the voltage increment rate for activity classification, it results in high classification error. Instead, with a larger $T_C \geq$ 4s, the voltage increment among different activities are more distinctive, which results in better classification accuracy. In the following section, we will evaluate the performance of \SystemName using our prototype.

\section{System Evaluation}
\label{section:system_evaluation_keh}
%In this section, we present in detail the set of experiments we have conducted to evaluate the overall performance of \SystemName and demonstrate the feasibility of this novel concept for daily activity recognition. %cIn particular, we aim to evaluate the sensing accuracy of \SystemName given different subjects, energy accumulation times.%; and (2) the energy harvesting performance of our prototype, i.e., how much energy can be harvested from different activities.

%In particular, we aim to evaluate the following two major aspects of \SystemName: (1) the sensing accuracy of \SystemName given different subjects, and energy accumulation times; and (2) the energy harvesting performance of our prototype, i.e., how much energy can be harvested from different activities.

\subsection{Experimental Setup and Data Collection}

\begin{figure}[]
	\centering
	\subfigure[Subject doing different activities.]{
		\includegraphics[scale=0.76]{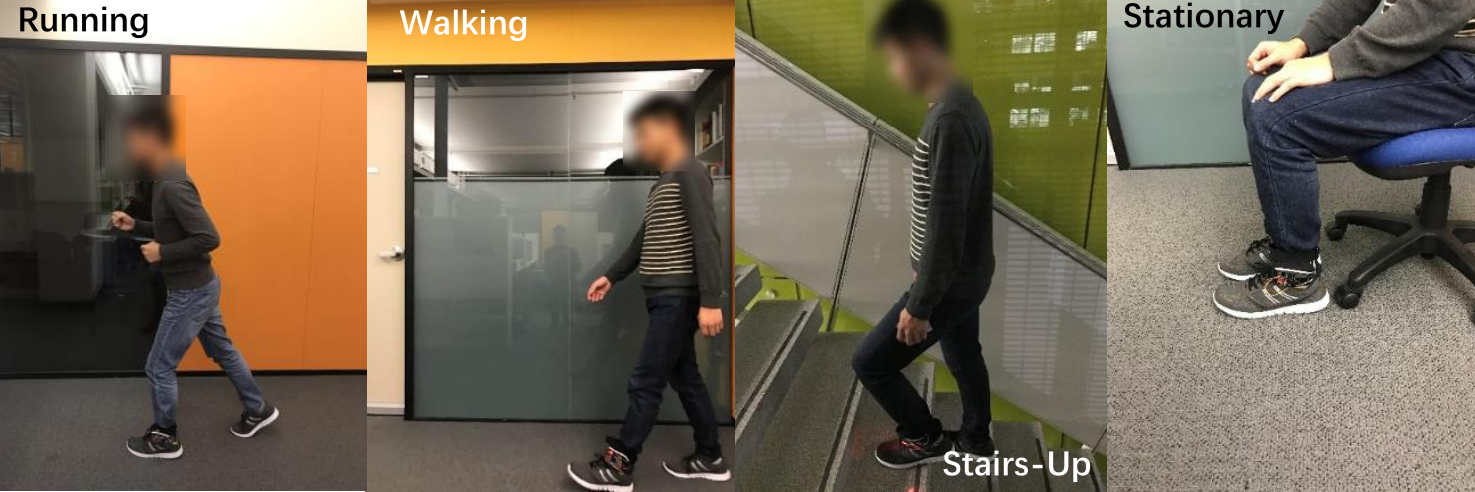}
		\label{fig:pehdifferentday}}
	\subfigure[Indoor environment.]{
		\includegraphics[width=2.2in]{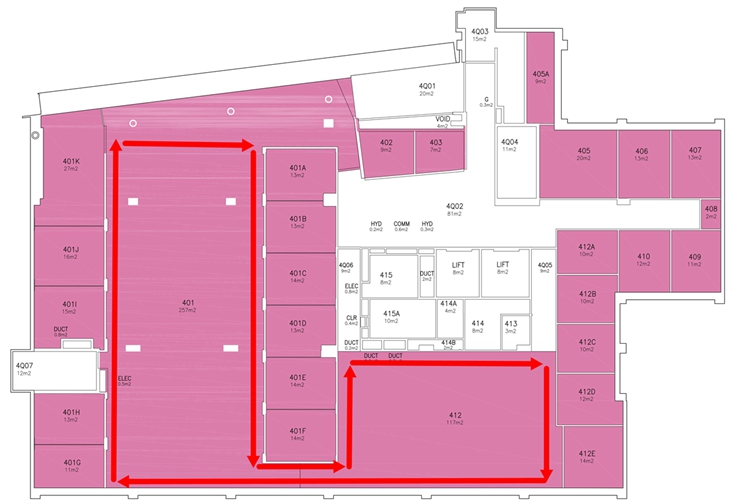}
		\label{fig:indoor}
	}
	\subfigure[Outdoor environment.]{
		\includegraphics[width=2.2in]{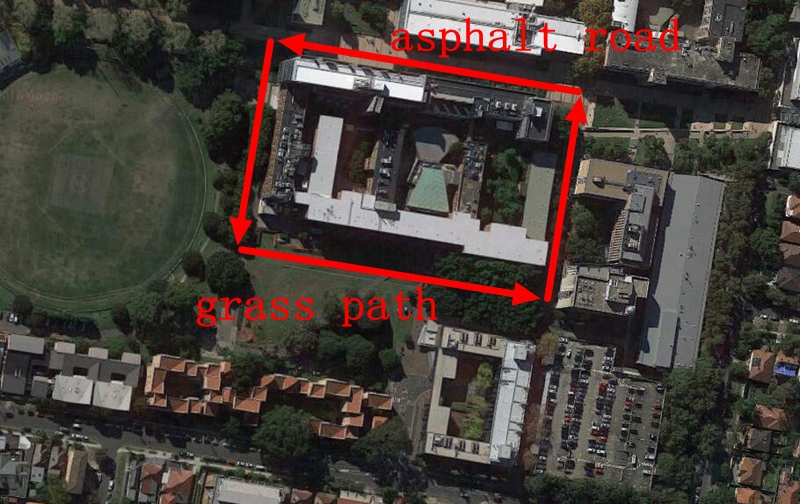}
		\label{fig:outdoor}
	}
	\caption{The illustration of data collection.}
	\label{Fig:data_collection}
	%\vspace{-0.1in}
\end{figure}

The subjects were asked to wear our energy-harvesting embedded shoe during the data collection. We prepared shoes with different sizes to meet the requirement of our subjects. The prototype system is attached to the subject's ankle (as shown in Figure~\ref{Fig:PEHhardware}). The dataset we used to evaluate the proposed system is collected from 10 healthy subjects who volunteered to do the experiments in our lab\footnote{Ethical approval for carrying out this experiment has been granted by the corresponding organization (Ethical Approval Number: HC15888).}. The subjects are diverse in gender (8 males and 2 females), age (range from 24 to 30), weight (from 55 to 75Kg), and height (from 168 to 183cm). We considered five different activities, including: walking (WALK), running (RUN), ascending stairs (SU), descending stairs (SD), and stationary (ST, i.e., sitting or standing).  Then, the subjects were asked to perform the activities normally in their own way without any specific instruction. As illustrated in Figure~\ref{Fig:data_collection}, for activities such as running and walking, they are performed in both indoor and outdoor environments to capture the influence of different terrains. For ascending and descending stairs, we conducted data collection in two building environments with different styles of stairs. For all the five activities, each volunteer participated at least two data collection sessions for both indoor and outdoor environments. For walking, running, and stationary, each session lasts at least 20 seconds, whereas, for ascending/descending stairs (i.e., the slope and steps of the stairs are different), each session may last 6 to 10 seconds depending on the number of steps and the walking speed of the subject. For each of the five activities, we have collected at least four sessions of samples from each of the 10 subjects. In total, we have 210 sessions of data. During the data collection, an Arduino Uno is used for data logging. The voltage of both front and rear capacitors are sampled and stored on the SD card at 100Hz sampling rate for offline data analysis.

Recall our discussions in Section~\ref{sec:activity_sensing_capacitor_voltage} that \SystemName leverages the increasing rate $r$ over the last accumulation time of $T_C$ for activity recognition. Following Equation~\ref{E4}, we calculate the $r$ of different activities with different $T_C$ for all the 10 subjects using our dataset. The estimated set of $r$ are then used as input for activity classification.

\subsection{Activity Recognition Performance}

The evaluation is carried out in WEKA\footnote{WEKA: http://www.cs.waikato.ac.nz/ml/weka/.} using 10-folds cross validation with 10 repetitions for each test. Four typical machine learning algorithms are used: the C4.5 decision tree algorithm (C4.5)~\cite{quinlan2014c4}, IBk K-Nearest Neighbor classifier~\cite{aha1991instance}, Naive Bayes with kernel estimation~\cite{john1995estimating}, and RandomForest~\cite{Breiman2001}. Those classifiers have been widely used in activity recognition and shown to be effective with high accuracy~\cite{bulling2014tutorial}. The parameters of the classifiers are optimized using the CVParameterSelection algorithm~\cite{Kohavi1995Thesis}. We evaluate \SystemName performance in terms of activity recognition accuracy (i.e., True Positive Rate). In the following, we evaluate the performance of \SystemName given different subjects, positions of PEH, energy accumulation times, and classifiers.

\subsubsection{\textbf{Recognition Accuracy vs. Subject}}
%The first thing we are interested in is the impact of subject difference on the recognition accuracy. 
Intuitively, given the diversity in subject's gender, weight, and height, the foot strike pressures applied on the energy harvesters differ in the way subjects perform the activity. In the following, we consider RandomForest as the classifier and fix the accumulation window $T_C=6s$ to investigate the impact of the subject difference on the classification accuracy. %The effect of accumulation window and classifier on the accuracy will be discussed shortly.

%One evidence of such impact has been shown in Table~\ref{tab:average_harvesting_power}, which indicates that the average harvesting power differs by subjects. Again, as discussed in Section~\ref{sec:activity_sensing_capacitor_voltage}, \SystemName leverages the capacitor voltage increment to estimate the energy harvesting rate for activity recognition, the subject difference may also affect the system accuracy.   

\begin{figure}[]
	\centering
	\includegraphics[width=3.6in]{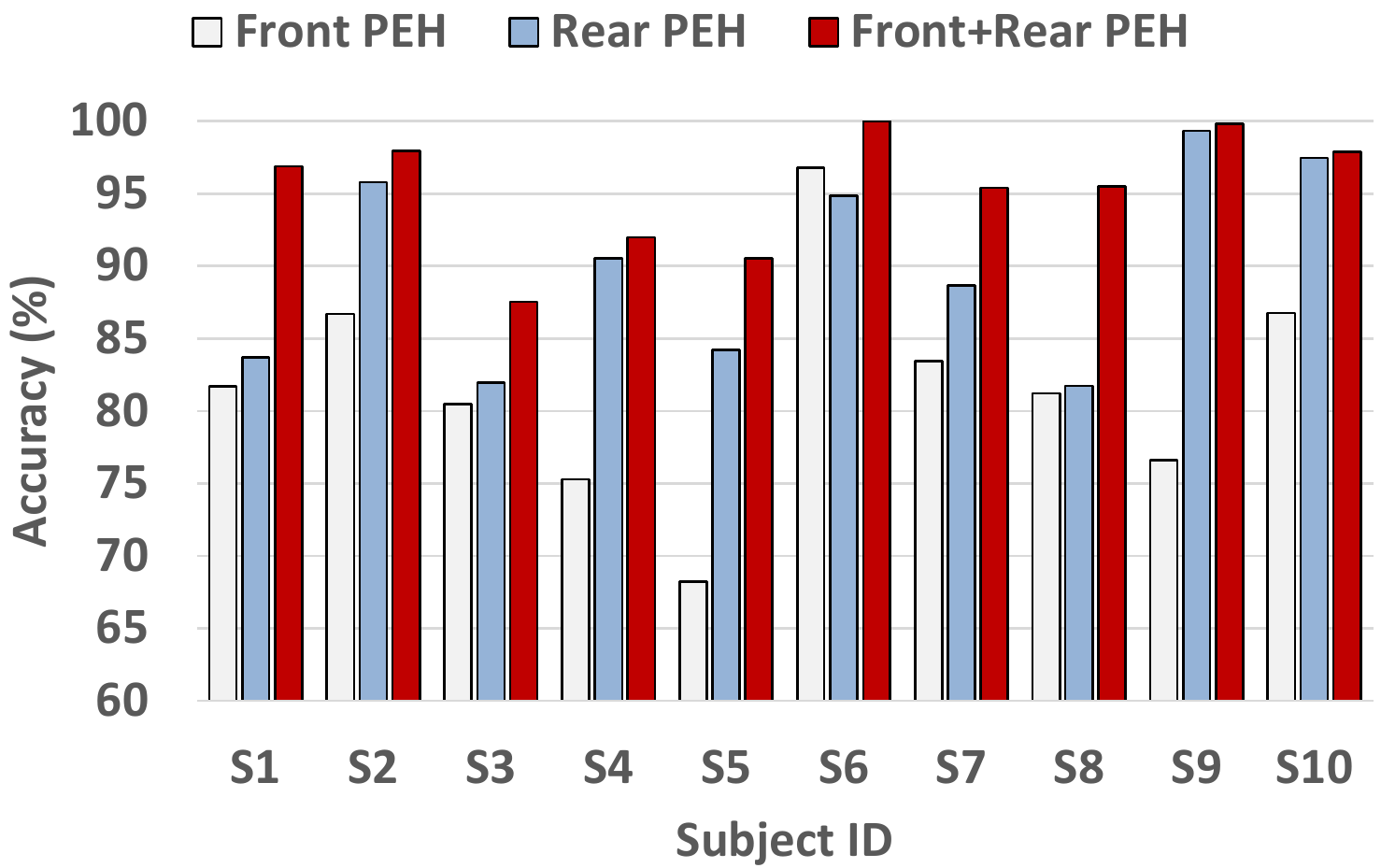}
	\caption{\SystemName recognition accuracy (in \%) achieved for all the ten subjects. The classifier is RandomForest and the accumulation windows is fixed as $T_C=6s$.} \label{Fig:Result_RF}
	%\vspace{-0.1in}
\end{figure}

\begin{figure*}[]
	\centering
	\subfigure[Subject 4, Front PEH.]{
		\includegraphics[scale=0.52]{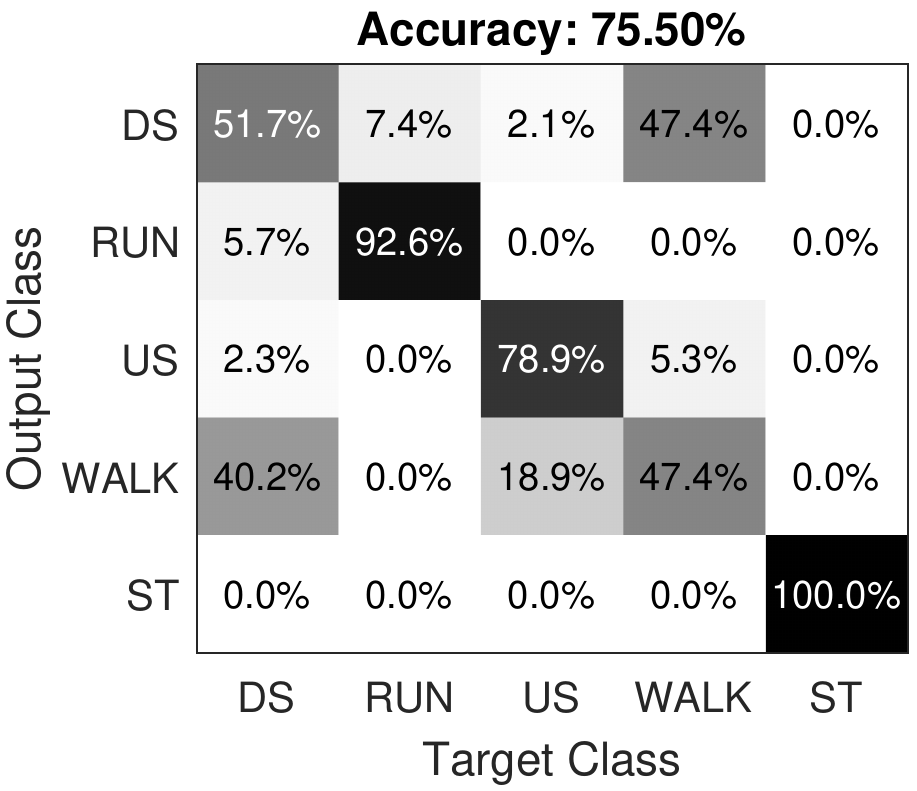}
		\label{fig:CM_S4_FRONT}
	}
	\subfigure[Subject 4, Rear PEH.]{
		\includegraphics[scale=0.52]{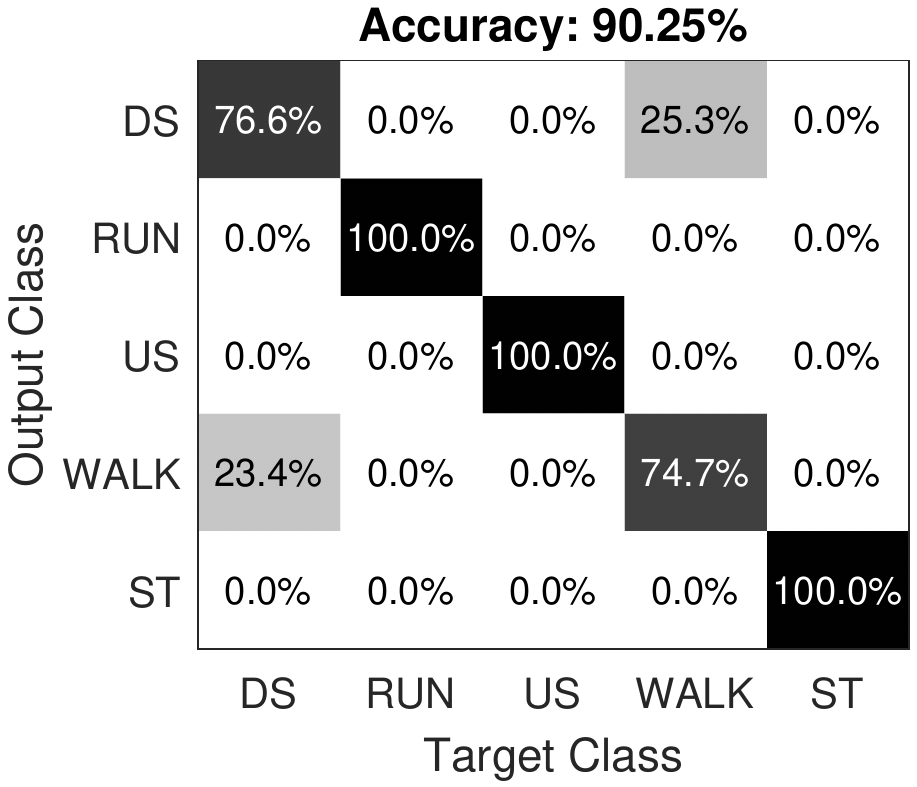}
		\label{fig:CM_S4_BACK}
	}
	\subfigure[Subject 6, Front PEH.]{
		\includegraphics[scale=0.52]{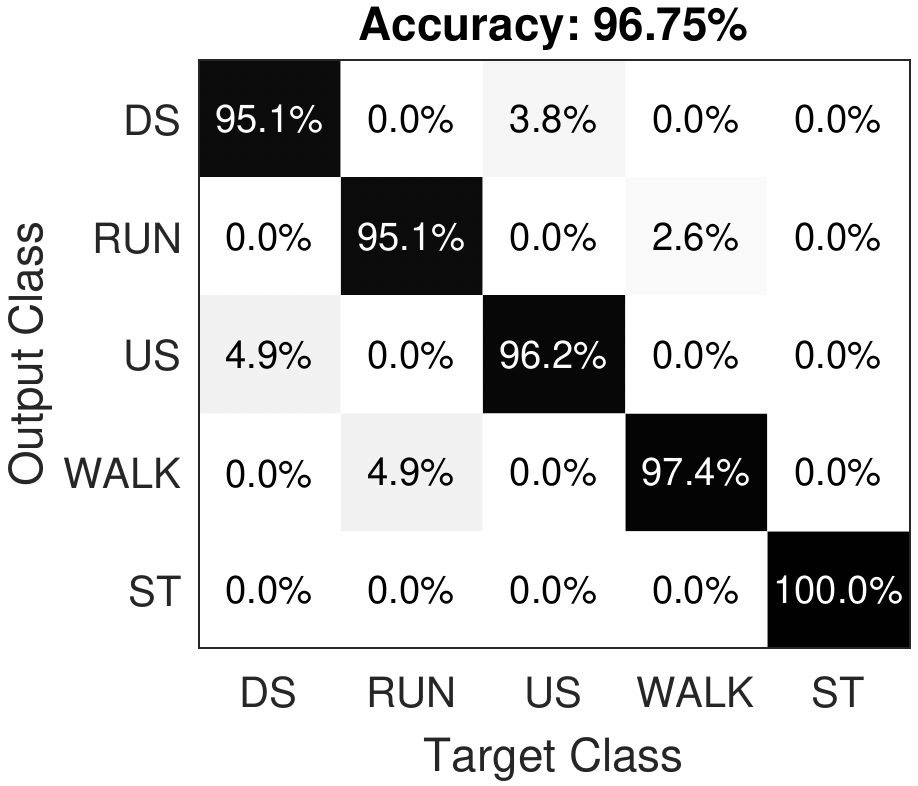}
		\label{fig:CM_S6_FRONT}
	}
	\caption{Confusion matrix of \SystemName with RandomForest classifier for Subject 4 and 6. The $T_C=6s$.}
	\label{Fig:Subject_CM}
	\vspace{-0.1in}
\end{figure*}

\begin{figure*}[]
	\centering
	\subfigure[Subject 4, Front PEH.]{
		\includegraphics[scale=0.5]{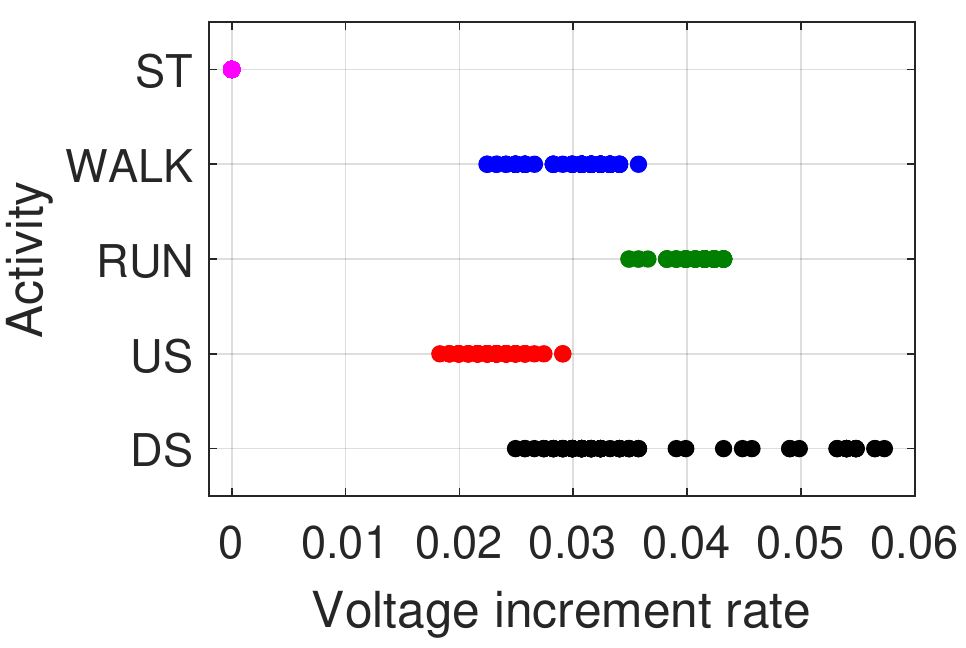}
		\label{fig:SC_S4_Front}
	}
	\subfigure[Subject 4, Rear PEH.]{
		\includegraphics[scale=0.5]{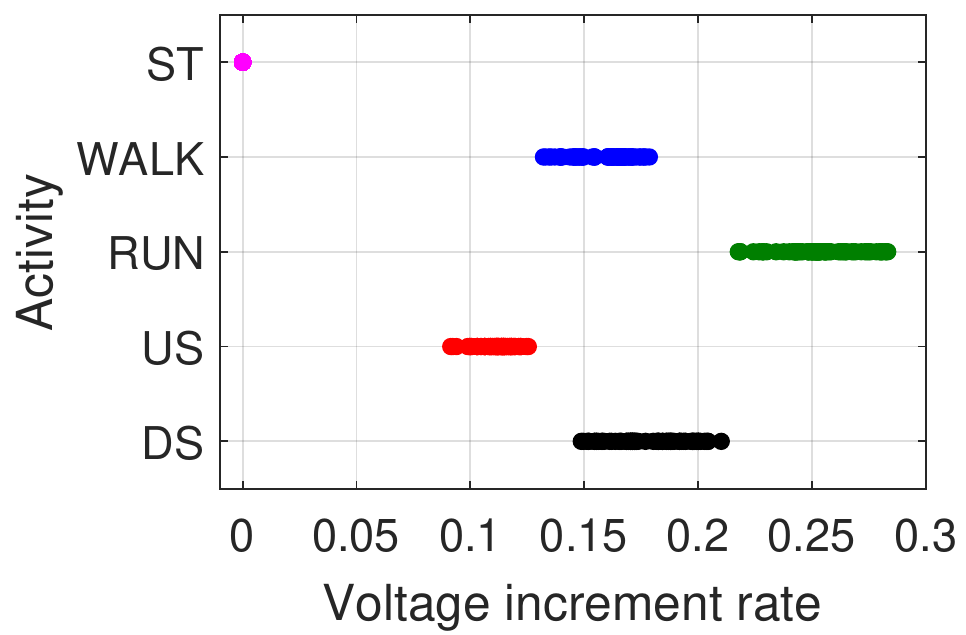}
		\label{fig:SC_S4_Back}
	}
	\subfigure[Subject 6, Front PEH.]{
		\includegraphics[scale=0.5]{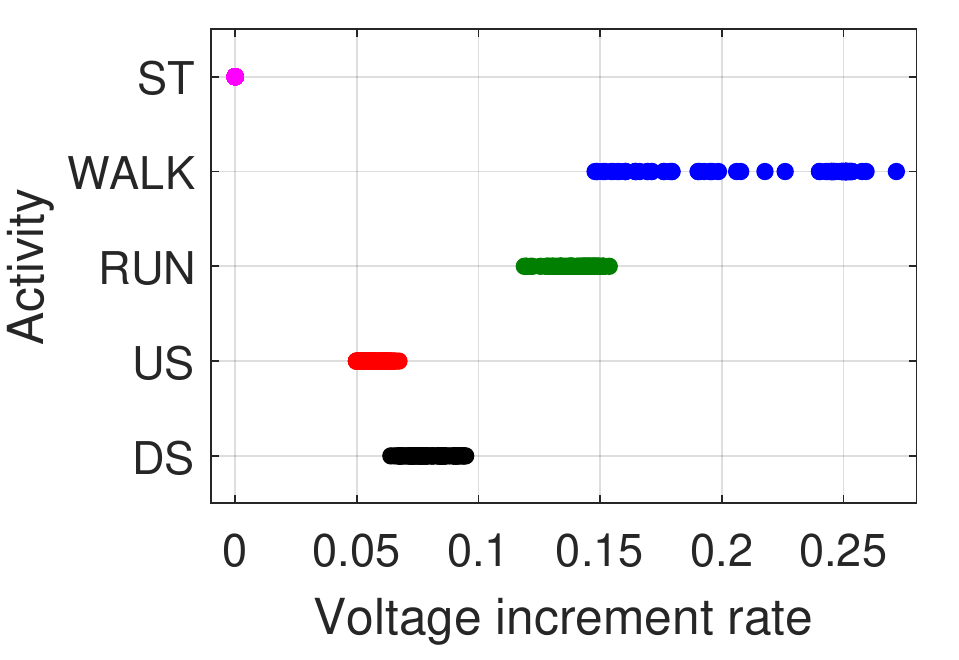}
		\label{fig:SC_S6_Front}
	}
	\caption{Scatter plots of the voltage increment rate of the capacitor when it is charged by the energy harvested from different activities. The Y-axis indicates the activity class, while the X-axis indicates the estimated capacitor voltage increment rate $r$ over an accumulation time $T_C$ of 6s. Each dot indicates an estimated instance of $r$ for a given activity.}
	\label{Fig:Subject_ST}
	\vspace{-0.1in}
\end{figure*}

Figure~\ref{Fig:Result_RF} exhibits the achieved accuracy for all the 10 subjects. As expected, the accuracy varies with subject. For example, Subject 4 achieves 75.50\% of accuracy using the voltage samples from the Front PEH as signal, whereas, Subject 6 achieves a much higher accuracy of 96.75\% using the same PEH. The corresponding confusion matrix are given in Figure~\ref{Fig:Subject_CM}. In addition, the scatter plots in Figure~\ref{Fig:Subject_ST} visualize all the instances of $r$ given different activities. \\%when the capacitor is charged by the power harvested from different activities. 

As shown in Figure~\ref{fig:CM_S4_FRONT}, the confusion matrix indicates that the major error happens in the classification between `WALK \& DS' , and `WALK \& DS'. This results from the high similarity in the voltage increment rate $r$ when Subject 4 is conducting those activities. As shown in Figure~\ref{fig:SC_S4_Front}, we can observe a high overlapping in $r$ between the activity `WALK', `US', and `DS'. That means, the voltage increment rate of the capacitor when it is charged by the energy harvested from those activities are very similar for Subject 4. This results in a very low accuracy of 75.5\%. Different from Subject 4, Subject 6 performs those activities in a different way. As shown in Figure~\ref{fig:SC_S6_Front}, the voltage increment rate of the capacitor when powered by those activities are more diverse. We can notice a very limited number of samples are overlapping among those activities. This results in a much higher classification accuracy as exhibited in the confusion matrix given in Figure~\ref{fig:CM_S6_FRONT}. This result suggests that \SystemName performs differently among subjects. We should take the subject difference into account when tuning the classification algorithm. In this regard, we conduct all the following experiments in a subject-dependent manner. 

\subsubsection{\textbf{Recognition Accuracy vs. Different PEH}}

In the following, we examine the accuracy achieved by using the signal from different PEHs, and investigate the possibility of using signal fusing from two PEHs to increase the accuracy. 

The results given in Figure~\ref{Fig:Result_RF} indicate that by using the capacitor voltage from the Rear PEH we can achieve a higher classification accuracy for most of the subjects (except Subject 6). Taking Subject 4 as an example again. We can notice from Figure~\ref{fig:SC_S4_Back} that the voltage increment rate are more separated among different activities for the Rear PEH when comparing to that of the Front PEH shown in Figure~\ref{fig:SC_S4_Front}. This observation matches with the confusion matrix given in Figure~\ref{fig:CM_S4_BACK} that only a small fraction of confusions happen between class `WALK' and `DS'. 

\begin{figure}[]
	\centering
	%	\subfigure[Scatter plot of the voltage increments.]{
	%		\includegraphics[width=1.95in]{Figures/S1_SCATTER_NEW.pdf}
	%		\label{Fig:S1_SCATTER}
	%	}
	\subfigure[Front PEH.]{
		\includegraphics[scale=0.52]{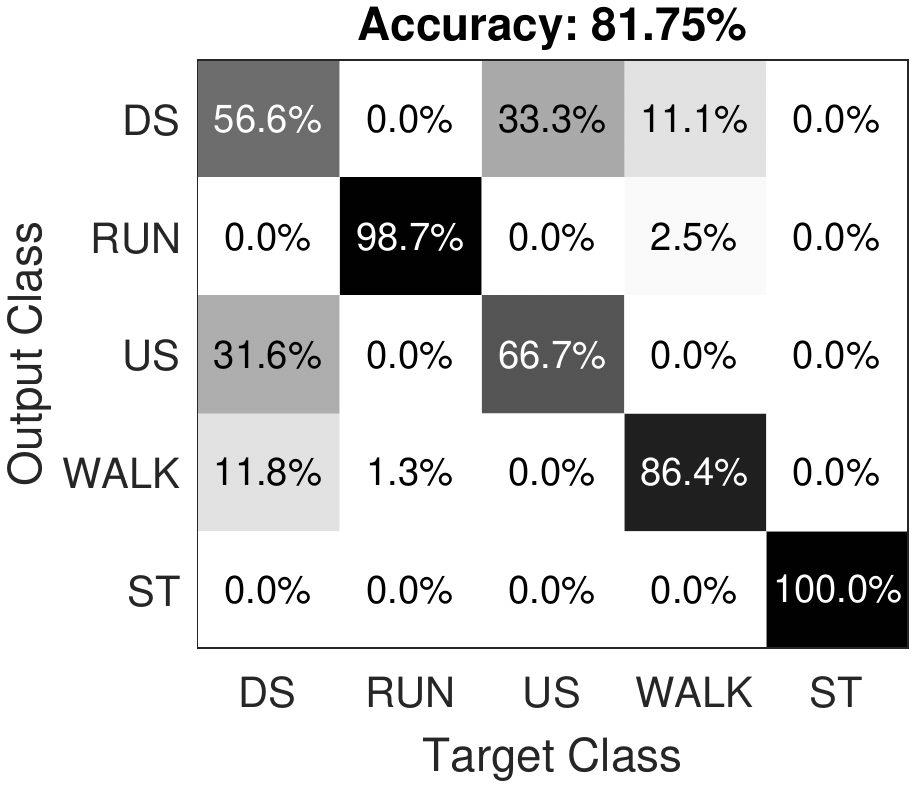}
		\label{fig:CM_S1_FRONT}
	}
	\subfigure[Rear PEH.]{
		\includegraphics[scale=0.52]{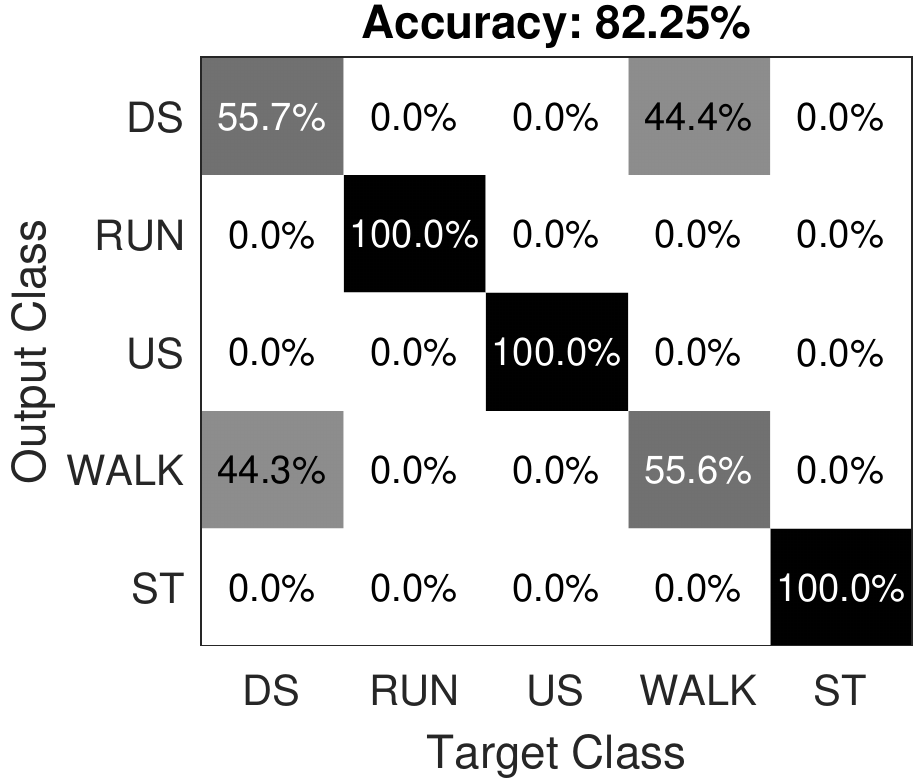}
		\label{fig:CM_S1_BACK}
	}
	\subfigure[Front + Rear PEHs.]{
		\includegraphics[scale=0.52]{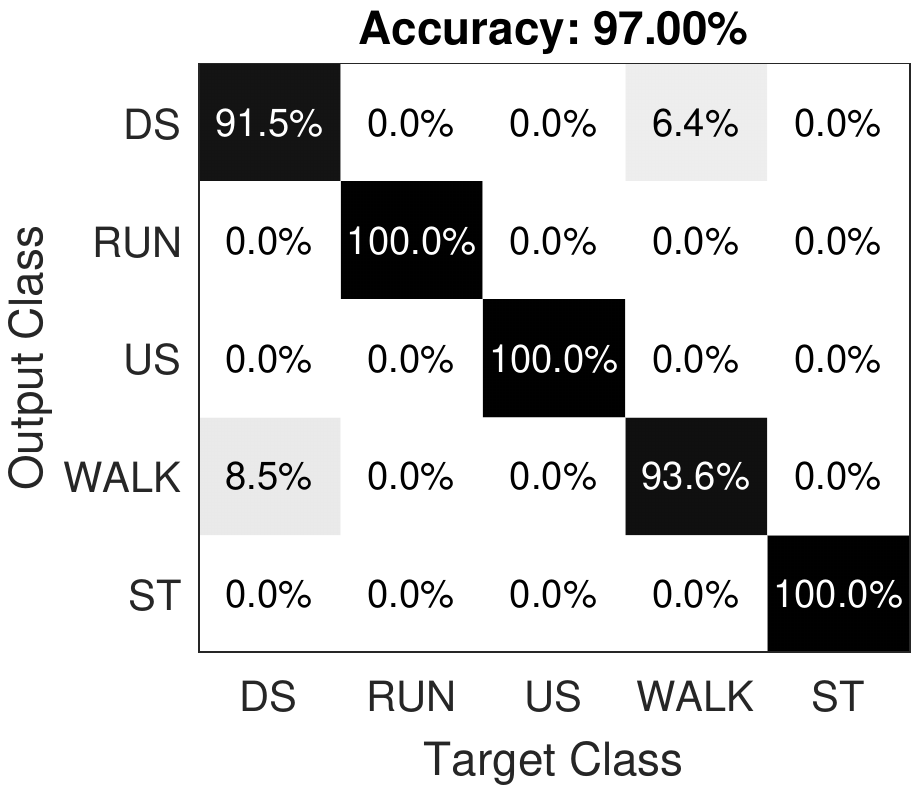}
		\label{fig:CM_S1_ALL}
	}
	\caption{The confusion matrix for Front, Rear, and Front+Rear PEH, respectively. The $T_C=6s$.}
	\label{Fig:Subject1_CM}
	\vspace{-0.1in}
\end{figure}

\begin{figure}[]
	\centering
	\includegraphics[width=3.in]{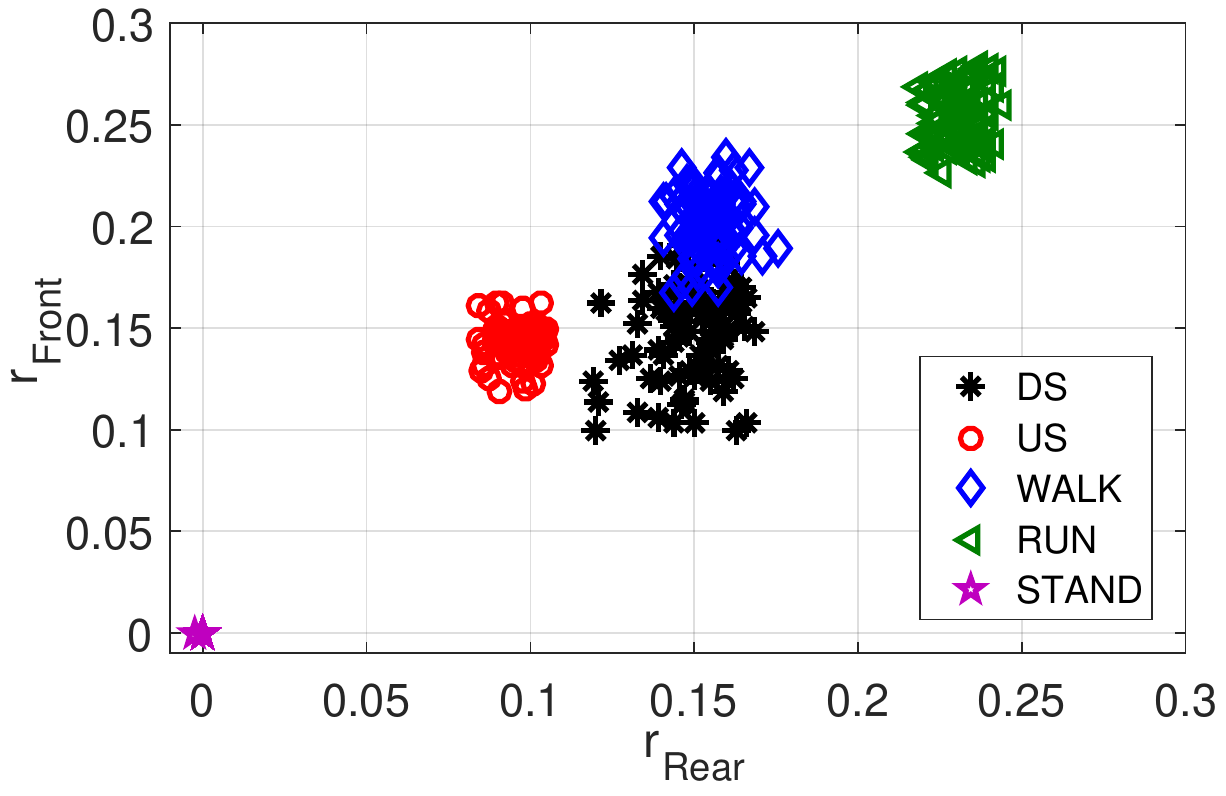}
	\caption{The scatter plot exhibits the samples of the two-dimensional vector, $<r_{Rear}, r_{Front}>$, when Subject 1 is doing different activities. The X and Y-axis indicates the capacitor voltage increment rate $r$ of the Rear and Front PEH, respectively.} \label{Fig:S1_SCATTER}
	\vspace{-0.1in}
\end{figure}

Moreover, \SystemName achieves better performance by fusing the signal from the two PEHs. That is, a two-dimensional vector, $<r_{Rear}, r_{Front}>$, is used as the input for classification, in which $r_{Rear}$ and $r_{Rear}$ refers to the capacitor voltage increment rate from the Rear and Front PEH, respectively. As an example, the scatter plot in Figure~\ref{Fig:S1_SCATTER} exhibits samples of $r_{Rear}$ and $r_{Rear}$ when Subject 1 is doing different activities. We can notice that, the $r_{Front}$ of activity `DS' and `US' are very similar to each other (as shown in Figure~\ref{Fig:S1_SCATTER}, considering $r_{Front}$ in the Y-axis only, most samples of `DS' and `US' fall in the same range from 0.1 to 0.16). That means, when Subject 1 conducts the activities `DS' and `US', the amounts of energy generated by the Front PEH from those two activities are very similar, which makes it hard to differentiate those two activities by using $r_{Front}$ only. As shown in Figure~\ref{fig:CM_S1_FRONT}, it results in high classification error between `DS' and `US'. Similar results apply to the Rear PEH as well. In Figure~\ref{Fig:S1_SCATTER}, if only consider the $r_{Rear}$ value in X-axis for classification, we can notice a high similarity between activity `DS' and `WALK' in $r_{Rear}$. Again, this results in high classification error as shown in Figure~\ref{fig:CM_S1_BACK}. 

However, as shown in Figure~\ref{fig:CM_S1_ALL}, after fusing the signal of Front and Rear PEHs, the confusions between those classes are significantly resolved. Intuitively, as exhibited in Figure~\ref{Fig:S1_SCATTER} and \ref{fig:CM_S1_ALL}, after fusing the signal from both Front and Rear PEHs and applying the two-dimensional voltage vector for classification, only a small fraction of samples are misclassified between `WALK' and `DS'.

\subsubsection{\textbf{Recognition Accuracy vs. Accumulation Time}}

\begin{figure}[]
	\centering
	\includegraphics[scale=0.68]{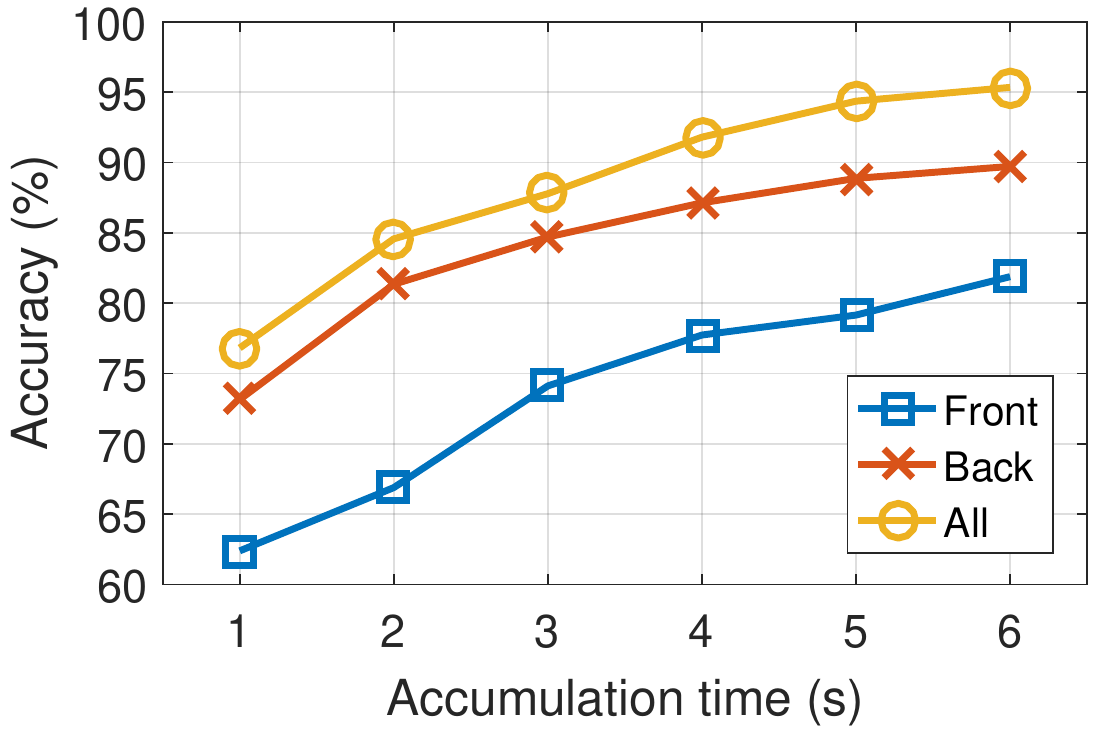}
	\caption{The accuracy (in \%) achieved by \SystemName  given different accumulation window $T_C$. The results are averaged across the 10 subjects. The classifier is RandomForest.} \label{Fig:Result_AVG_TIME}
	\vspace{-0.1in}
\end{figure}

Now, we investigate the impact of accumulation time $T_C$ on the recognition accuracy. Figure~\ref{Fig:Result_AVG_TIME} exhibits the achieved accuracy given different $T_C$. The classifier used in this experiment is RandomForest. For a particular $T_C$, the reported results are the averaged accuracy across all the 10 subjects. We can clearly observe that, regardless of the signal, the accuracy increases with $T_C$. Intuitively, as shown previously in Figure~\ref{Fig:capacitor_charging_example}, a larger accumulation window leads to a more distinctive difference in the capacitor voltage. Thus, a large $T_C$ is preferable in improving the classification accuracy. However, the sojourn time for a subject in performing a specific activity is short, and transitions between activities may occur in the middle of $T_C$. Therefore, $T_C$ should not be set too large that exceeds the activity sojourn time. As reported in~\cite{yan2012energy}, for activities such as walking, running, and standing/sitting, the sojourn time is at least 1 to 2 minutes. For activities such as ascending/descending stairs, the sojourn time is much shorter, but still longer than 5 seconds for over 99.9\% of the time. During our data collection, we have noticed that, for ascending/descending a 10-steps stair, the sojourn time is usually within 6 to 10 seconds depending on the subject's speed. Therefore, as a trade-off between the system performance and robustness, we recommend the maximum value of $T_C$ for \SystemName to be configured to 5 seconds. As shown in Figure~\ref{Fig:Result_AVG_TIME}, after fusing the signal from Front and Back PEHs, \SystemName is able to achieve 95\% of accuracy with $T_C=5s$. 

\subsubsection{\textbf{Recognition Accuracy v.s. Classifier}}
%\begin{figure}
%	\centering
%	\includegraphics[scale=0.62]{Figures/result_class_avg.pdf}
%	\caption{The accuracy (in \%) achieved by \SystemName  with different classifiers given different accumulation window $T_C$. The results are the averaged accuracy across the ten subjects.} \label{Fig:Result_Classifier}
%	%\vspace{-0.15in}
%\end{figure}

\begin{table}[]
	\centering
	\caption{The accuracy (in \%) achieved by \SystemName  with different classifiers given accumulation window $T_C=5s$. The results are averaged accuracy across the 10 subjects.}
	\label{table:resualts_different_classifiers}
	\resizebox{3.9in}{!}{
		\begin{tabular}{|l|c|c|c|c|}
			\hline
			\multirow{2}{*}{} & \multicolumn{4}{c|}{\textbf{Classifier}}   \\ \cline{2-5} 
			& Naive Bayes & IBK   & J48   & RandomForest \\ \hline \hline
			\textbf{Front PEH}    & 78.95       & 78.11 & 79.62 & 78.36        \\ \hline
			\textbf{Back PEH}     & 89.46       & 88.63 & 89.11 & 88.88        \\ \hline
			\textbf{Front+Back}      & 95.08       & 93.83 & 94.57 & 94.87        \\ \hline
		\end{tabular}
	}
\vspace{-0.05in}
\end{table}

Lastly, we analyze the performance of \SystemName with different classifiers. Table~\ref{table:resualts_different_classifiers} exhibits the accuracy of \SystemName with different classifiers given $T_C=5s$. The results are averaged across all the 10 subjects. We can observe that, all the four examined classifiers can achieve over 93\% of accuracy after fusing the signal from the two PEHs. This performance is comparable to that achieved by conventional motion sensor-based systems~\cite{lara2013}. An interesting observation is that \SystemName exhibits no bias on the selection of classifier, as all the four classifiers achieved similar classification results. 

\section{Energy Consumption Analysis}
\label{section:energy_consumption}

High energy consumption is the major roadblock for the pervasive use of wearable technology~\cite{seneviratne2017survey}. In this section, we will conduct an extensive power consumption profiling of off-the-shelf wearable activity recognition systems to investigate the superiority of \SystemName in energy saving. We will demonstrate that, by dramatically reduce the sampling frequency down to 0.2Hz, \SystemName allows the MCU to stay in the energy-saving low-power mode for extended periods of time comparing to the state-of-the-art KEH transducer-based system~\cite{khalifa2017harke}. Consequently, we will show that \SystemName can reduce the system power consumption by several factors.

\subsection{Setup for Energy Consumption Analysis}

We use an off-the-shelf Texas Instrument SensorTag\footnote{SensorTag: http://www.ti.com/ww/en/wireless\_connectivity/sensortag/.} as the target device, which is embedded with the ultra-low power ARM Cortex-M3 MCU that is specifically designed for today's energy-efficient wearable devices\footnote{Mainstream wearable devices such as FitBit are using ARM Cortex-M3: see \url{https://www.ifixit.com/Teardown/Fitbit+Flex+Teardown/16050}.}. The SensorTag is running the Contiki 3.0 operating system\footnote{Contiki OS: http://www.contiki-os.org/.} which duty-cycles the MCU to save energy. Moreover, all unnecessary components, including the onboard ADC, SPI bus, and the on board accelerometers are powered-off when it is not sampling. We are interested in the power consumption of SensorTag in data sampling (i.e., either in sampling the capacitor voltage or the AC voltage signal from KEH transducer), and the power consumption in data transmission. The average power consumption and time requirement for each sampling and transmission events are measured by using the built-in function in the Agilent DSO3202A oscilloscope. In the following, we present our measurement results of those parts in order. 

\subsection{Power Consumption in Sampling}

\begin{figure}
	\centering
	\includegraphics[scale=0.4]{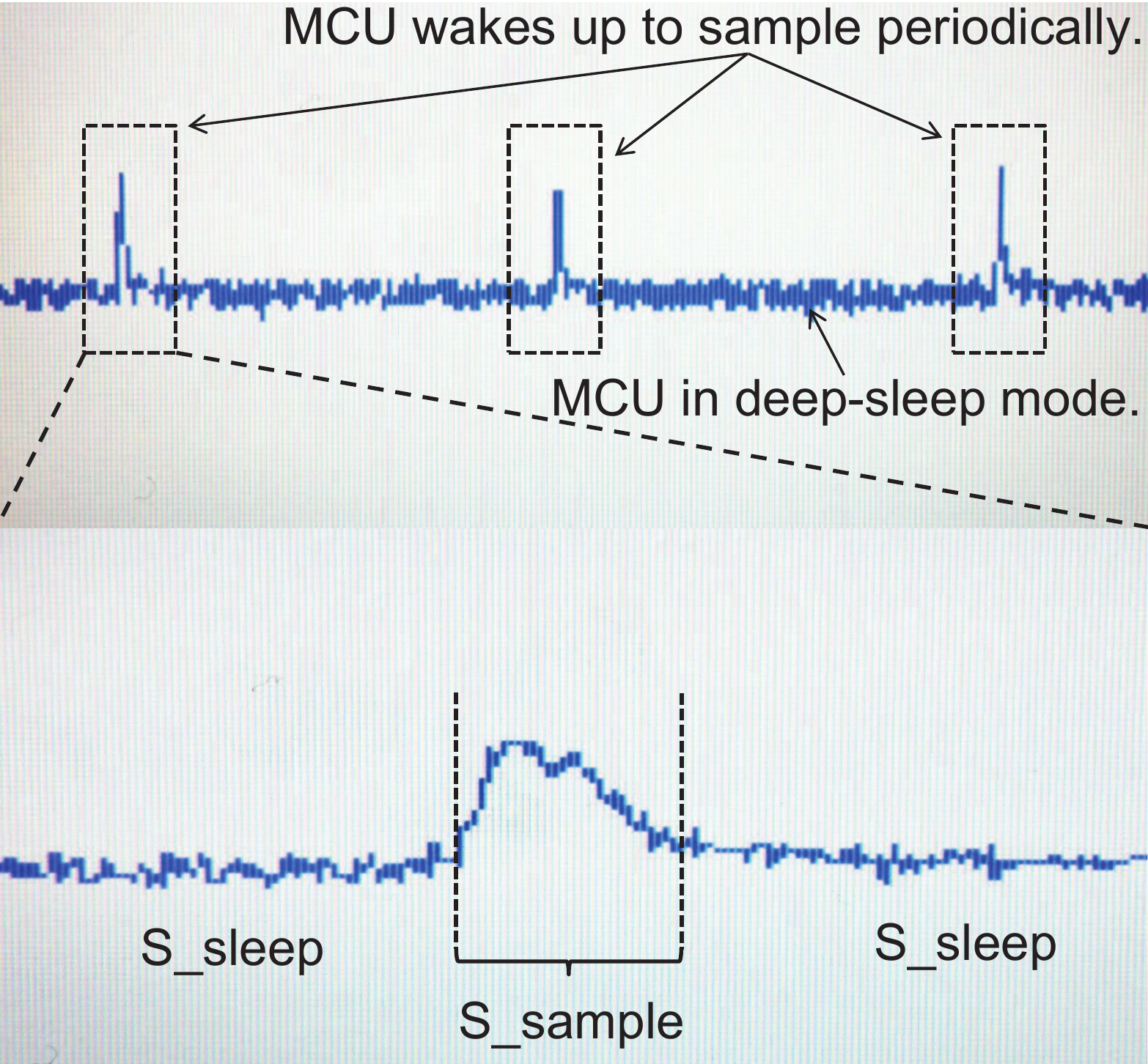}
	\caption{Profiling of voltage sampling.} \label{Fig:Energy_Measurement_KEH}
	%\vspace{-0.1in}
\end{figure}

First, we investigate the power consumption in data sampling. In the measurement, both capacitor voltage and KEH transducer signal are simpled through the on board ADC of SensorTag. The sampling frequency of MCU is configured as 25Hz to meet the requirement of KEH transducer-based sensing system~\cite{khalifa2017harke}. Figure~\ref{Fig:Energy_Measurement_KEH} presents an oscilloscope trace when MCU sampling the signal from ADC periodically. As shown, MCU is triggered by the timer to sample periodically. It takes approximately $0.6ms$ for the MCU to complete a single voltage sampling event (i.e., state $S_{sample}$ shown in Figure~\ref{Fig:Energy_Measurement_KEH}). After that, MCU turns back into the deep sleep mode (i.e., LPM3 in Contiki OS) to save power. The average power consumption of the system for a single ADC sampling is 480$\mu$W, and the baseline system power consumption when MCU is in the deep-sleep-mode is only 6$\mu$W. The details are summarized in Table~\ref{table:keh_power_consumption}. 

\begin{table}[]
	\centering
	\caption{States of MCU in sampling the ADC signal.}
	\label{table:keh_power_consumption}
	\resizebox{4in}{!}{
		\begin{tabular}{|l|c|c|l|}
			\hline
			\textbf{State} & \begin{tabular}[c]{@{}l@{}}\textbf{Time}\\ (ms)\end{tabular} & \begin{tabular}[c]{@{}l@{}}\textbf{Power}\\ ($\mu$W)\end{tabular} & \multicolumn{1}{c|}{\textbf{Description}} \\ \hline \hline
			$S_{sample}$ & 0.6 & 480 & MCU wakes up to sample ADC signal. \\
			$S_{sleep}$ & null & 6 & MCU in deep-sleep mode. \\ \hline
		\end{tabular}
	}
	%\vspace{-0.1in}
\end{table}

In general, for the duty-cycled activity sensing system, the average power consumption in data sampling, $P_{sense}$, can be obtained by the following equation:
\begin{equation}
P_{sense} = 
\begin{cases} \frac{T_S\times n}{1000} P_{sample} + (1-\frac{T_S\times n}{1000}) P_{sleep}  & \text{if $0 \leq n \leq \frac{1000}{T_S}$,}\\
P_{sample} & \text{if $\frac{1000}{T_S} < n$.}
\end{cases}
\label{Power_equation}
\end{equation} 
where, $P_{sample}$ is the average power consumption of the system during the sampling event, and $P_{sleep}$ is the average power consumption when the MCU is in deep-sleep mode (with all the other system components power-off). $n$ is the sampling frequency, and $T_S$ is the duration of time (in milli-second) required by a single sampling event. Based on the measurement results given in Table~\ref{table:keh_power_consumption}, we can have the average power consumption for KEH voltage sampling event, $P_{sample}$, equals to 480$\mu$W with a duration, $T_S$, equals to 0.6ms. The power consumption when MCU in deep-sleep mode, $P_{sleep}$, is 6$\mu$W. For KEH transducer-based system, given different application scenarios, a sampling frequency of 25Hz-50Hz is required to achieve good accuracy for human activity recognition~\cite{khalifa2017harke,weitao2016ndss,lan2016transportation}. Therefore, given the minimum required sampling frequency of 25Hz, following Equation~\ref{Power_equation} we can obtain the power consumption in data sampling for KEH transducer-based system equals to 13.11$\mu$W. On the other hand, as demonstrated in Section~\ref{section:system_evaluation_keh}, to achieve an overall classification accuracy of 90\%, \SystemName need only to sample the ADC signal once every 5s. Thus, given Equation~\ref{Power_equation}, the power consumption in data sampling for \SystemName is only 6.06$\mu$W. 

\begin{table}[]
	\centering
	\caption{The power consumption ($\mu$W) in data sampling.}
	\resizebox{3.4in}{!}{
		\begin{tabular}{|l|c|c|}
			\hline
			& \textbf{KEH transducer-25Hz} & \textbf{CapSense-0.2Hz} \\ \hline \hline
			Sensing & 7.11 & 0.06 \\ 
			MCU Sleep & 6 & 6 \\ 
			Overall & 13.11 & 6.06 \\ \hline
		\end{tabular}
	}
	\label{talbe:power_compare}
	%\vspace{-0.1in}
\end{table}

The results are compared in Table~\ref{talbe:power_compare}. As shown, \textbf{\SystemName is able to save 54\% of the overall power consumed by the transducer-based system in data sampling}. We can also notice that, for \SystemName, the main energy expenditure is the MCU Sleep (i.e., the unavoidable power consumption of the system when MCU is in the deep-sleep mode), which consumes 99\% (6$\mu$W over 6.06$\mu$W) of the overall sampling power consumption. Fortunately, with the rapid development of energy-efficient micro-controllers, we can expect the power consumption of MCU Sleep can be further reduced. For instance, the STM32L4 Series of MCUs\footnote{STM32L4 Series: http://http://www.st.com/en/microcontrollers/stm32l4-series.html.} consume only 1.35$\mu W$ in its sleep mode, it can help \SystemName achieving an ultra-low system power consumption of 1.41$\mu W$.

\subsection{Power Consumption for Data Transmission}
\begin{figure}[]
	\centering
	\includegraphics[scale=0.5]{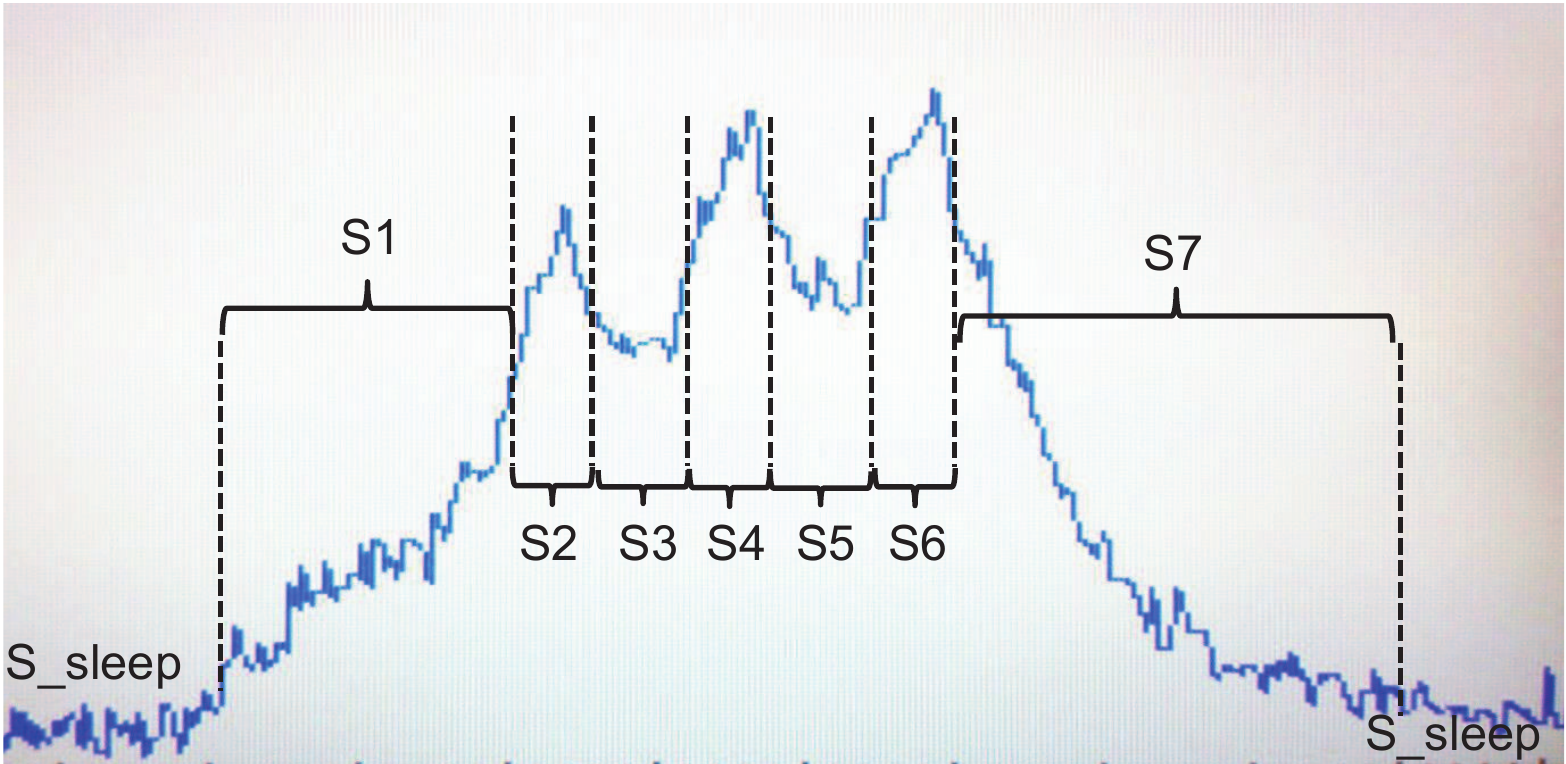}
	\caption{Profiling of BLE broadcasting event.} \label{Fig:Energy_Measurement_BLE}
	%\vspace{-0.1in}
\end{figure}

In the following, we investigate the power consumption in wireless data transmission using the BLE beacons. We programmed the Contiki OS to wake-up the CC2650 wireless MCU periodically and transmits a BLE beacon packet, which broadcasts three times on three separate channels (repetition improves reliability of broadcasting). The transmitted beacon packets are all 19 Bytes (all for protocol payloads. The details for each BLE broadcasting event are visualized in Figure~\ref{Fig:Energy_Measurement_BLE}, and summarized in Table~\ref{table:ble_power_consumption}. Note that, the transmission time is depending on the packet size. The time stated in Table~\ref{table:ble_power_consumption} (0.28ms for state S2, S4, and S6) is the minimum required time to transmit the 19 Bytes packet per channel. For every additional Byte\footnote{According the protocol, up to 28 Bytes of data could be added to the 19 Bytes payloads per packet.} to be transmitted, 8$\mu$s time needs to be added to the total transmission time. 

\begin{table}[]
	\centering
	\caption{States of BLE broadcasting event.}
	\label{table:ble_power_consumption}
	\resizebox{4.5in}{!}{
		\begin{tabular}{|l|c|c|l|}
			\hline
			\textbf{State} & \begin{tabular}[c]{@{}l@{}}\textbf{Time}\\ (ms)\end{tabular} & \begin{tabular}[c]{@{}l@{}}\textbf{Power}\\ (uW)\end{tabular} & \multicolumn{1}{c|}{\textbf{Description}} \\ \hline \hline
			S1 & 1.12 & 1008 & Radio setup.\\
			S2, S4, S6 & 0.28 & 3990 & Radio transmits a beacon packet of 19 Bytes.\\
			S3, S5 & 0.30 & 2460 & Transition between transmissions.\\
			S7 & 1.72 & 744 & Post-processing before sleep.\\
			S\_sleep & null & 6 & Radio off; MCU in deep-sleep mode.\\ \hline
		\end{tabular}
	}
\end{table}

\begin{table}[] \small
	\centering
	\caption{Summary of data transmission power consumption.}
	\label{talbe:transmission_power}
	\resizebox{2.6in}{!}{
		\begin{tabular}{|l|c|c|}
			\hline
			& \multicolumn{1}{l|}{\textbf{KEH-25Hz}} & \multicolumn{1}{l|}{\textbf{CapSense-0.2Hz}} \\ \hline \hline
			Power & 3.129$m$W & 1.716$m$W\\
			Time & 24.25$ms$ & 4.3$ms$\\ 
			Energy & 75.89$\mu J$ & 7.43$\mu J$\\\hline
	\end{tabular}}
	%\vspace{-0.1in}
\end{table}	

For transducer-based system with a sampling frequency of 25Hz, it has $25Hz\times 5s = 125$ voltage samples (2 Bytes for each 12-Bits ADC reading, and 250 Bytes in total) to be transmitted per channel once every five seconds. Given the maximum additional data can be added to each beacon packet is 28 Bytes, this requires $\lceil{\frac{250}{28}}\rceil =9$ packets to be transmitted per channel. As a result, for transducer-based system, it consumes 75.89$\mu J$\footnote{Obtained by: $1.12ms \times 1.008mW + 1.72ms \times 0.744mW+27 \times (0.28ms+0.008ms \times 28) \times 3.99mW + 26 \times 0.3m s \times 2.46mW = 75.89\mu J.$} to transmit the 9 packets on three different channels. The average power consumption is 3.129$m$W with time duration of 24.25$ms$. On the other hand, for \SystemName, it has only one voltage sample to be transmitted once every five seconds (in total, 2 Bytes), results in one packet to be transmitted per channel. The total energy consumption for \SystemName is only 7.43$\mu J$\footnote{Obtained by: $1.12ms \times 1.008mW + 1.72ms \times 0.744mW+3 \times (0.28ms+0.008ms \times 2) \times 3.99mW + 2 \times 0.3ms \times 2.46mW = 7.43\mu J.$}. The average power consumption is 1.716$m$W with time duration of 4.3$ms$.	The results are compared in Table~\ref{talbe:transmission_power}. As shown, \SystemName is able to \textbf{save over 90.2\% of the energy consumption in data transmission}. Clearly, for KEH transducer-based systems, the radio has to stay for a longer period of time to transmit more sampling data. Although, different transmission approaches (data aggregation and feature selection) can be applied to reduce the amount of data to be transmitted~\cite{lara2013}, and thus, reduce the transmission power consumption. However, additional on-board computations for those mechanisms may still introduce inevitable power consumption. 

Combining the power consumption in data sampling and transmission together, the overall system power consumption for KEH transducer-based system is 28.15$\mu W$\footnote{Obtained by:$(\frac{13.11\mu W \times 5sec + 3.129mW\times 24.25ms}{5sec+24.25ms})=28.15\mu W.$}, whereas, the overall system power consumption for \SystemName is only 7.53$\mu$W\footnote{Obtained by:$(\frac{6.06\mu W \times 5sec + 1.716mW\times 4.3ms}{5sec+4.3ms})=7.53\mu$W.}. This means that \textbf{\SystemName is able to save 73\% of the overall system power consumption of state-of-the-art KEH transducer-based system}.

\section{Related Work}
\label{section:related_work}

In this section, we review existing works in developing energy-efficient mobile sensing system. We first review the efforts in building insole-based self-powered wearable system. Then, we introduce some recent efforts in utilizing KEH-transducer as the motion sensor for energy-efficient sensing. Lastly, we review some works in reducing sampling-induced power consumption by finding the minimum required sampling frequency for conventional motion sensor-based activity recognition systems.

\subsection{Insole-based Energy Harvesting System}
With recent advances in energy harvesting hardware, researchers are now turning to kinetic energy harvesting as a viable source of power to extend battery life or even replace the batteries altogether in wearable devices \cite{paradiso2005energy,mitcheson2008energy}. Some wearable KEH products are already appearing in the market, such as AMPY wearable motion charger~\cite{AMPY}, SEQUENT self-charing smartwatch~\cite{SEQUENT}, and SOLEPOWER energy harvesting shoe~\cite{SOLEPOWER}, showing signs of promising future for this technology. In the context of wearable shoes, insole-based kinetic energy harvesting is widely regarded as the most popular solution to achieve self-power given the high harvesting efficiency from human walking~\cite{antaki1995gait}. The history of building shoe-based self-powered wearable devices starts from the late nineties. In an earlier work of Antaki~et al.~\cite{antaki1995gait}, the authors discovered that the ground reaction forces associated with the heel strike and toe-off phases of the gait can generate the largest amount of energy during human walking. In this study, a piezoelectric array-based EH shoe has been built to generate electric energy from human gait. Similarly, in the work of Kymissis~et al.~\cite{kymissis1998parasitic}, a piezoelectric generator is placed inside the shoe and can generate 1.1-1.8mW average power during walking. They have proved that the generated energy is able to power the RFID transmitter to broadcast signal periodically. More recently, in~\cite{meier2014piezoelectric}, a shoe-mounted energy harvesting system has been developed for podiatric analysis. A piezoelectric energy harvester was leveraged to generate 10-20$\mu$J energy per waling step. A pressure sensor and passive footstrike sensor were utilized to analysis human gaits. Another example of self-powered shoe is given by Huang~et al.~\cite{huang2015battery}. In their prototype consists a pair of shoes, an accelerometer and Bluetooth wireless communication unit are powered separately by the energy harvested from each of the two feet, and coordinated by ambient backscatter. Such that, the accelerometer can sense the activity of the user, while the Bluetooth can transmit the sensing results to the smartphone. Different from the aforementioned efforts that focus either on maximizing the amount of harvesting energy, or optimizing the wearable system to achieve self-power, the focus and contribution of our work is to investigate the feasibility of utilizing the capacitor voltage to achieve daily activity recognition while dramatically reduce the required sampling frequency. To the best of our knowledge, this has not been studied in the current literature yet.

\subsection{KEH-transducer based Context Sensing}

Thanks to the existing efforts in kinetic-powered wearable systems, some studies in the literature start to apply KEH transducer as a low power vibration sensor. The motivation behind this idea is to further reduce the energy consumption in powering conventional motion sensor, e.g., accelerometer. In~\cite{khalifa2017harke}, Khalifa et al. proposed the idea of using the power signal generated by a KEH device for human activities recognition. The proposed system can achieve $83\%$ of accuracy for classifying different daily activities. In~\cite{lan2016transportation}, the authors investigate the feasibility of using KEH as the sensor for transportation mode detection. Similarly, in~\cite{kalantarian2015monitoring}, a piezoelectric transducer-based wearable necklace has been design for food-intake monitoring. The proposed system achieves over 80\% of accuracy in distinguishing food categories. In~\cite{blank2016ball}, Blank et al. proposed a ball impact localization system using a piezoelectric embedded table tennis racket. More recently, Xu et al.~\cite{weitao2016ndss} proposed an authentication system which utilizes the AC voltage signal to authenticate the user based on gait analysis. The proposed system can achieve an recognition accuracy of $95\%$ when five gait cycles are used. In~\cite{lan2017veh,lan2018hidden}, the authors proposed the use of KEH-transducer as an energy-efficient receiver for acoustic communication. 

\subsection{Reducing Sampling Frequency}
For both KEH-transducer based and conventional motion sensor based sensing systems, the energy consumption in sensing is proportional to the sampling frequency. Thus, a large volume of works in the literature focused on reducing the sampling rates to save the energy~\cite{bulling2014tutorial,ghasemzadeh2015power,yan2012energy,qi2013adasense} to improve the system energy efficiency. For instance, in~\cite{krause2005trading}, Krause et al., studied the trade-off between the system power consumption and classification accuracy by using a smartwatch wearable device. They demonstrated that the lifetime of the device can be extended by selecting the optimal sampling strategy without accuracy losing. Similar results are presented in~\cite{yan2012energy}, in which the authors pointed out that there is a trade-off between sampling frequency and classification accuracy, and introduced the A3R algorithm which adapts the sampling frequency and classification features in real-time based on the activity type. In addition, by leveraging the temporal-sparsity of human activity, researchers have also proposed the use of compressive sensing theory to reduce sampling frequency~\cite{chou2009energy,li2013compressed,qi2016gazelle}. Instead of reducing the sampling frequency to the level of tens of Hz, in this work, we introduce \SystemName to bring the sampling frequency down to 0.2 Hz.c

\section{Conclusion}
\label{section:conclusion}

In this paper, we present \SystemName, a novel activity sensing scheme for KEH-powered wearable devices. By simply using the voltage readings of the energy harvesting capacitor at 0.2Hz, \SystemName is able to daily activities with 95\% accuracy, and reduce system power consumption by 73\%. The current work is a first step in capacitor-based sensing for KEH-powered IoTs. As such, it can be extended in many directions. First of all, as the current hardware prototype is quite cumbersome, another direction for future work is to design the prototype with a smaller form-factor, and provide detailed user study on the practical user experience of this device. Second, as our hardware can harvest energy from different user activities, we will investigate ways to utilize the harvested energy to power our system, thus making it battery-free. Lastly, in addition to human activity recognition, we would like to explore the feasibility of \SystemName in different scenarios and considering different types of energy harvesters, such as, the monitoring of appliance usage in an smart-home environment. For instance, by leveraging capacitors powered by the thermo and solar energy harvester, it may be possible to detect the usage of hot-water and indoor lights. 

% Bibliography
\bibliographystyle{ACM-Reference-Format-Journals}
\bibliography{mybib}

\medskip
\end{document}